\documentclass[journal]{IEEEtran}
\IEEEoverridecommandlockouts
\usepackage{cite}
\usepackage{subfigure}
\usepackage[font=footnotesize]{caption}
\usepackage{multirow}
\usepackage{bm}
\usepackage{hyperref}
\usepackage{algorithm}
\usepackage{amsmath,amssymb,amsfonts}
\usepackage{graphicx}
\usepackage{textcomp}
\usepackage{xcolor}
\usepackage{threeparttable}
\usepackage{tablefootnote}
\usepackage{makecell}
\usepackage[]{algpseudocode}
\algtext*{EndFunction}
\algtext*{EndWhile}
\algtext*{EndProcedure}
\algtext*{EndFor}
\algtext*{EndIf}
\def\BibTeX{{\rm B\kern-.05em{\sc i\kern-.025em b}\kern-.08em
    T\kern-.1667em\lower.7ex\hbox{E}\kern-.125emX}}

\begin{document}

\title{H-VGRAE: A Hierarchical Stochastic Spatial-Temporal Embedding Method for Robust Anomaly Detection in Dynamic Networks}
\author{\IEEEauthorblockN{Chenming Yang\IEEEauthorrefmark{2}, Liang Zhou\IEEEauthorrefmark{2}, Hui Wen\IEEEauthorrefmark{1}, Zhiheng Zhou\IEEEauthorrefmark{2} and Yue Wu\IEEEauthorrefmark{1}}

\IEEEauthorblockA{\IEEEauthorrefmark{2}National Key Laboratory of Science and Technology on Communications, \\
\IEEEauthorrefmark{1}Department of Computer Science and Engineering,\\
University of Electronic Science and Technology of China, Chengdu, P. R. China\\
Email: \{chenmingyang, huiwen\}@std.uestc.edu.cn, \{lzhou, zhzhou, ywu\}@uestc.edu.cn}}

\maketitle
\begin{abstract}
Detecting anomalous edges and nodes in dynamic networks is critical in various areas, such as social media, computer networks, and so on.
Recent approaches leverage network embedding technique to learn how to generate node representations for normal training samples and detect anomalies deviated from normal patterns.
However, most existing network embedding approaches learn deterministic node representations, which are sensitive to fluctuations of the topology and attributes due to the high flexibility and stochasticity of dynamic networks.
In this paper, a stochastic neural network, named by Hierarchical Variational Graph Recurrent Autoencoder (H-VGRAE), is proposed to detect anomalies in dynamic networks by the learned robust node representations in the form of random variables.
H-VGRAE is a semi-supervised model to capture normal patterns in training set by maximizing the likelihood of the adjacency matrix and node attributes via variational inference.
The encoder of the H-VGRAE encodes hierarchical spatial-temporal information of topology and node attribute into multi-layer conditional random variables, and then the decoder reconstructs the dynamic network based on the latent random variables.
For a new observation of the dynamic network, the reconstruction probabilities of edges and node attributes can be obtained from the trained H-VGRAE, and those with low reconstruction probabilities are declared as anomalous.
Comparing with existing methods, H-VGRAE has three main advantages: 1) H-VGRAE learns robust node representations through stochasticity modeling and the extraction of multi-scale spatial-temporal features; 2) H-VGRAE can be extended to deep structure with the increase of the dynamic network scale; 3) the anomalous edge and node can be located and interpreted from the probabilistic perspective.
Extensive experiments on four real-world datasets demonstrate the outperformance of H-VGRAE on anomaly detection in dynamic networks compared with state-of-the-art competitors.
\end{abstract}




\section{Introduction}
Anomaly detection for networks (graphs) has been an active topic in various research areas, such as social media, computer security, and e-commerce \cite{aggarwal2015outlier, akoglu2015graph, ranshous2015anomaly}.
Numerous studies have been conducted on static networks in the past two decades by capturing deviations of structural patterns and show great success in practical applications \cite{akoglu2015graph}.
However, with the rapid growth of data and users, the update of network topology and node attributes becomes increasingly frequent.
Typical examples include IP-IP network traffic logs, item buying and comments in e-commerce sites, system components logs in cyber-physical systems, and so on.
The patterns of anomalies are thus time-varying and difficult to capture if merely applying methods from static networks \cite{aggarwal2011outlier}.
Due to the complex spatial-temporal (ST) patterns, identifying anomalous edges and nodes in such dynamic networks remains a challenging problem.

A general framework to detect anomalies in dynamic networks is to extract the features that contribute to the anomaly detection, and then compute anomaly scores based on the features.
The critical problem is how to extract anomaly-related features.
Some studies predefine the anomalies by considering structural changes, such as sudden changes in connectivity, node-degree distribution, and so on \cite{eswaran2018spotlight,yoon2019fast}.
However, the anomalous patterns are much more than sudden structural changes and could be very complicated in practical scenarios.
It is thus necessary to implicitly explore the techniques that can extract anomaly-related features implicitly in the data-driven mode.

Recently, a novel feature selection technique called network embedding has shown promising results on extracting features of networks, especially on preserving structural information in the low-dimensional representations \cite{li2019mad, cui2018survey}.
The learned features can be utilized in downstream tasks, such as node classification, link prediction, as well as anomaly detection, without having to define them explicitly.
Due to the anomaly diversities and the lack of labels for training, the anomaly detection methods based on network embedding has to work in the semi-supervised mode \cite{chalapathy2019deep}.
The model needs to be trained to capture normal patterns from the training set and detects anomalies according to the deviations from the learned normal patterns.

In this paper, we also focus on detecting anomalies in dynamic networks based on network embedding, similar to NetWalk \cite{yu2018netwalk} and AddGraph \cite{zheng2019addgraph}.
However, since dynamic networks often interact with the environment and human operators to show complex evolving patterns, their behaviors can exhibit both stochasticity and strong spatial-temporal (ST) correlations.
The network embedding technique used in NetWalk \cite{yu2018netwalk} and AddGraph \cite{zheng2019addgraph} can only generate deterministic latent representation, which cannot model the stochasticity well and could lead to lack of robustness in anomaly detection when facing rapid dynamics.
Moreover, limited with the utilized neural network models in NetWalk \cite{yu2018netwalk} and AddGraph \cite{zheng2019addgraph}, the ST information is still relatively coarse and could also cause the models sensitive to tiny changes.
Therefore, the goal of this paper is to learn robust node representations to capture normal patterns of dynamic networks by modeling stochasticity and complex ST dependencies.
Ideally, with only a few anomalous edges, the node representations would be dominant by the normal pattern, and thus the anomalous edges and nodes would have low reconstruction probabilities.
There are two major challenges to achieve our goal.

The first challenge is how to combine stochasticity and ST dependency together in the latent representation and enhance the representation capability.
The combination of stochasticity and temporal dependence has been conducted in \cite{chung2015recurrent} by unifying Recurrent Neural Network (RNN) and variational autoencoder (VAE) \cite{VAE}, and be extended to graph domain in Variational Graph Recurrent Neural Network (VGRNN) \cite{hajiramezanali2019variational} by additionally extracting spatial features using graph neural networks (GNN).
Unfortunately, constrained by GNN and RNN, which only consider the one-hop neighbors in node representation and temporal correlations in adjacent timestamps, the ST features extracted by VGRNN may lead the detector sensitive to local and short-term changes.
And the capability of the latent representations is limited due to the single layer of stochastic variables in VGRNN .

To strengthen the latent representations, we propose a Hierarchical Variational Graph Recurrent autoencoder (H-VGRAE), which is scalable to deep structure and considers multi-scale ST information.
H-VGRAE encodes stochasticity, multi-scale ST and content features into the hierarchical stochastic latent representation.
H-VGRAE has three major advantages in architecture:
a) Through hierarchical ST blocks connected with multiple stochastic layers, H-VGRAE can capture multi-scale ST patterns, which has been shown to be useful to strengthen robustness in anomaly detection tasks \cite{chung2016hierarchical}.
b) By sharing information about the conditional stochastic latent variables between the encoder and the decoder, H-VGRAE can model more complex stochasticity by stacking multiple layers of stochastic variables, alleviating the difficulty to train deep conditional stochastic layers.
c) H-VGRAE can model non-Gaussian distributions by employing Normalizing Flow \cite{rezende2015variational} in multiple correlated stochastic layers.
Benefited from these advantages, H-VGRAE is capable of modeling complex patterns of normal behaviors in the dynamic networks and provides rich information to be adjust to high variabilities.

Given the learned node representations, the second challenge is how to detect anomalies accurately and interpretably.
Considering the semi-supervised mode, the H-VGRAE is essentially maximizing the evidence lower bound (ELBO) of the likelihood of the adjacency matrix and node attributes in the normal training set.
In the online detection stage, the H-VGRAE captures the normal parts of an anomalous snapshot in the latent representations and reconstructs the normal edges and node attributes with high probability.
The multi-scale features and stochasticity modeling enhance the robustness of the latent representations.
The remaining ones with low reconstruction probabilities are detected as anomalies, which allows estimating the contribution of each dimension of node attributes.

Moreover, most of deterministic embedding based methods rely on single anomaly detection mechanism, such as clustering in latent representation \cite{yu2018netwalk} and specifically designed anomaly score \cite{zheng2019addgraph}.
In H-VGRAE, the anomaly detection combines reconstruction and prediction by designing the prior to predict the latent variables based on historical states.
Benefited from prediction prior, H-VGRAE essentially detects anomalies in two hybrid perspectives: 1) the input that cannot be reconstructed well is an anomaly; 2) the input that deviates from the prediction is an anomaly.

To summarize, the main contributions of this paper are as follows:
\begin{itemize}
  \item We analyze the likelihood of adjacency matrices and node attributes in multiple snapshots and obtain the evidence lower bound of the likelihood.
  \item We propose a novel semi-supervised network embedding method, H-VGRAE, which can capture normal patterns in dynamic networks by maximizing the evidence lower bound.
  H-VGRAE can jointly utilize stochasticity, spatial-temporal information, and content information to empower the embeddings.
  \item We design multiple mechanisms to strengthen the robustness of the node representations learned by H-VGRAE, including the extraction of multi-scale ST features, the generalization of non-Gaussian random variables, the predictive prior, and the information-sharing mechanism between the inference of posterior and prior to train deep hierarchies of conditional stochastic layers.
  \item The anomalies are detected based on the conditional reconstruction probabilities.
  The level of anomalies can be compared in different dynamic networks.
  \item Through extensive experiments on four real-world dynamic network datasets, we show the effectiveness and superiority of H-VGRAE in terms of AUC score. In dynamic networks with different scales, H-VGRAE exhibits great robustness, with AUC higher than 0.75 even with 10\% anomalous samples in the most complex and attributed Github network.
\end{itemize}

\section{Related Work}
In this section, we briefly review network embedding techniques and existing anomaly detection methods for dynamic networks.
\subsection{Network Embedding}
Network embedding assigns nodes in a network to low-dimensional representations, which effectively preserves the network structure and support network inference \cite{cui2018survey}.
For dynamic networks, methods have been developed using various techniques such as matrix factorization \cite{zhu2016scalable,zhang2018timers}, random walk \cite{nguyen2018continuous,yu2018netwalk}, neural networks \cite{goyal2020dyngraph2vec, seo2018structured, hajiramezanali2019variational, 8365780}, and stochastic process \cite{zhou2018dynamic, trivedi2019dyrep}.
Most of these methods focus on sudden changes in either structural features or content features (from node attributes) to learn dynamic node representations \cite{yang2016revisiting}, and only a few of them model both kinds of changes simultaneously \cite{hajiramezanali2019variational, trivedi2019dyrep}.
Moreover, due to the simplification of temporal patterns, such as smoothness \cite{zhang2018timers} and single granularity \cite{hajiramezanali2019variational}, the capability of existing models is limited.

Recent developments \cite{chung2015recurrent} in speech sequence have shown that the stochasticity can be more precisely captured by stochastic variables with properly estimated probability distributions than deterministic variables.
Inspired by the stochastic models in sequence processing, a recent study VGRNN \cite{hajiramezanali2019variational} introduces stochasticity into dynamic network embedding to model the uncertainty of the latent representations and obtain state-of-the-art performance in link prediction task.
However, constrained by the stochastic RNN in their models, the layers of stochastic variables in VGRNN cannot be deepened and thus cannot capture multi-scale ST features well.

%
%

\subsection{Anomaly Detection in Dynamic Networks}
According to the utilization of network embedding, the anomaly detection methods in dynamic networks can be classified into two categories (for other classification criterion, see \cite{akoglu2015graph} for survey):
\begin{itemize}
  \item \emph{Anomaly detection without using network embedding}: Most of existing methods define anomalies as sudden changes of structural patterns. The structural patterns are captured through connectivity \cite{aggarwal2011outlier}, sketch \cite{ranshous2016scalable}, dense subgraphs \cite{sun2006beyond, eswaran2018spotlight, shin2017d}, and so on. Some works design approximation mechanisms of the structural patterns to achieve near real-time pattern updating \cite{eswaran2018spotlight, miz2019anomaly, yoon2019fast}.
  Then, the anomaly scores are defined according to the structural patterns and the anomalies are detected by setting a anomaly score threshold.
  \item \emph{Anomaly detection using network embedding}: Recently, some works begun to leverage network embedding techniques to implicitly define anomalous patterns in dynamic networks.
  In \cite{sricharan2014localizing}, a commute time distance is used to localize anomalous changes in dynamic graphs, while the method mainly focuses on sudden structural changes and cannot capture anomalies with complex temporal patterns.
  NetWalk \cite{yu2018netwalk} utilizes an autoencoder to generate low-dimensional latent representations of nodes based on random walks in each timestamp, and detect anomalies by clustering these node representations dynamically. AddGraph \cite{zheng2019addgraph} extends NetWalk by using a spatial-temporal neural network to capture more complex patterns in dynamic networks, and employing negative sampling to build an end-to-end semi-supervised learning model.
\end{itemize}

\section{Preliminaries}
\subsection{Problem Formulation}

\textbf{Dynamic network model.}
Let $T_{max}$ be the maximum timestamp.
A network stream with length $T_{max}$ can be represented as $\mathbb{G} = \{\mathcal{G}(t)\}_{t=1}^{T_{max}}$, where $\mathcal{G}(t)=(\mathcal{V}(t),\mathcal{E}(t))$ is the entire snapshot at timestamp $t$, and $\mathcal{V}(t)$ and $\mathcal{E}(t)$ are the sets of nodes and edges respectively.
For convenience, we denote the union of $\mathbb{G}$ by $G = (V,E)$, where $V = \bigcup\nolimits_{t = 1}^{T_{max}} \mathcal{V}(t)$ and $E = \bigcup\nolimits_{t = 1}^{T_{max}} \mathcal{E}(t)$, and $N = |V|$ can denote the maximum number of nodes in all the $T_{max}$ timestamps.
The topology of the network $\mathcal{G}(t)$ can be represented by an adjacency matrix $\mathbf{A}(t) = [A_{ij}(t)] \in \{0,1\}^{N \times N}$, where $A_{ij}(t)$ means there exists an edge from node $i$ to node $j$ at timestamp $t$.
In practical scenarios, the dynamic may also have attributes attached on each node, which can be denoted as $\mathbf{X}(t)=(\mathbf{x}_{1}(t),\ldots,\mathbf{x}_{N}(t)) \in \mathbb{R}^{N \times D}$, where $D$ denotes the dimension of node attributes that is constant across time.

\textbf{Problem statement.}
For dynamic network anomaly detection, the objective is to detect anomalous edges and nodes at any given timestamp $t$, i.e., in real time as $\mathcal{G}(t)$ occurs.
As in time series modeling, historical observations can be beneficial for understanding current data.
Therefore, the streams $\mathbf{A}({\leq}t)=\{\mathbf{A}(\tau)\}_{\tau =1}^{t}$ and $\mathbf{X}({\leq}t)=\{\mathbf{X}(\tau)\}_{\tau =1}^{t}$ are used to detect anomalies, instead of just using $\mathbf{A}(t)$ and $\mathbf{X}(t)$.

\subsection{Overall Structure}
\begin{figure}[t]
\centerline{\includegraphics[width=0.48\textwidth]{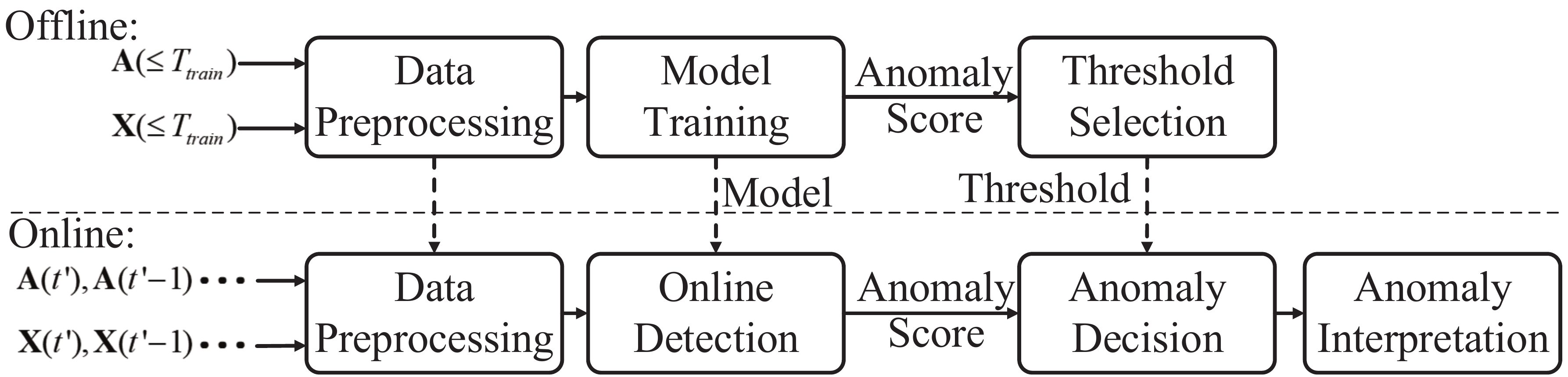}}
\caption{Overall Structure of semi-supevised anomaly detection.}
\label{Overall}
\end{figure}

As shown in Fig.\ref{Overall}, the overall structure of using H-VGRAE to detect anomalies is composed of two stages: offline model training and online anomaly detection.
In the offline stage, the proposed H-VGRAE is trained on normal training data $\mathbf{A}({\leq}T_{train})$ and $\mathbf{X}({\leq}T_{train})$ $(T_{train} \textless T_{max})$ to capture normal behaviors of the dynamic network by maximizing the evidence lower bound of the likelihood.
The reconstruction probabilities are used as anomaly scores.
Then the thresholds of anomaly scores to detect anomalous edges and nodes are set based on the anomaly detection performance on a validation dataset.
In the online stage, the trained H-VGRAE is used to calculate the anomaly scores of a new network observation $\mathcal{G}(t') (t' \textless T_{max})$ at timestamp $t'$, based on the all the current and previous online observations.
If the anomaly score of some edge/node in $\mathcal{G}(t')$ is higher than the threshold, it will be detected as an anomalous edge/node, otherwise, it is normal.
For the detected anomalous edges/nodes, the interpretations are made by the reconstruction probabilities of these edges/node attributes.

\subsection{Basics of Graph Autoencoder and Stochastic RNN}
\label{Stochastic RNN}
In network embedding, Graph Autoencoders (GAEs) \cite{GAE} are commonly used for their ability of modeling structure information.
Inheriting the encoder-decoder framework from the autoencoders, GAEs can learn low-dimensional latent representations of node features in the unsupervised mode.
The encoder is implemented with Graph Neural Network (GNN), which is able to fuse the content and structural information of a node with its neighbors.
Typical GNNs include GCN \cite{GCN}, GAT \cite{Veli2017} and GraphSAGE \cite{hamilton2017inductive}.
Then, the structure information is reconstructed by a link prediction decoder, like bilinear decoder.

Stochastic RNN is a variant of Bayesian model \cite{chung2015recurrent}, which contains a Variational Autoencoder (VAE) \cite{VAE} at every timestamp.
The latent random variable $\mathbf{z}(t)$ in Stochastic RNN are conditioned on the state variable $\mathbf{h}(t-1)$.
This condition helps the Stochastic RNN modeling highly variable temporal patterns of time series.
%
%
With input $\mathbf{x}({\leq}t)$, the Stochastic RNN have the inference stage $q_{\phi}(\mathbf{z}(t)|\mathbf{x}({\leq}t,\mathbf{z}(\textless{t}))$ and the generation stage $p_{\theta}(\mathbf{x}(t)|\mathbf{x}({\textless{t}}),\mathbf{z}({\leq}t))$
with a prior $p_{\text{prior}}(\mathbf{z}(t)|\mathbf{z}(\textless{t}))$.
For real-value inputs, all the three probability distributions can be set to diagonal Gaussian $\mathcal{N}(\boldsymbol{\mu}(t),\boldsymbol{\sigma}^{2}(t)\mathbf{I})$.
%
To model temporal dependency, there is a recurrence stage that updates the hidden state
\begin{equation}
  \mathbf{h}(t) = f(\varphi^{\mathbf{x}}(\mathbf{x}(t)),\varphi^{\mathbf{z}}(\mathbf{z}(t)),\mathbf{h}(t-1)),
\end{equation}
where $f(\cdot)$ is a deterministic non-linear transition function such as LSTM \cite{hochreiter1997long} or GRU \cite{chung2014empirical}; $\varphi^{\mathbf{x}}(\cdot)$ and $\varphi^{\mathbf{z}}(\cdot)$ are neural networks to extract features.
The parameters of the Stochastic RNN can be learned in a end-to-end mode, by maximizing the ELBO of likelihood:
\begin{equation}
  \label{SRNN}
  \begin{split}
  \mathcal{L} = &\sum\limits_{t=1}^{T} \mathbb{E}_{q_{\phi}(\mathbf{z}(t)|\mathbf{x}({\leq}t),\mathbf{z}(\textless t))}
  \log p_{\theta}(\mathbf{x}(t)|\mathbf{z}({\leq}t),\mathbf{x}(\textless t))\\
  &-\text{KL}(q_{\phi}(\mathbf{z}(t)|\mathbf{x}({\leq}t),\mathbf{z}(\textless t)) \parallel  p_{\text{prior}}(\mathbf{z}(t)|\mathbf{z}(\textless{t}) ))
\end{split}
\end{equation}

Due to the simple assumption of diagonal Gaussian, the KL divergence may vanish during training, leading to the generation stage trapped in sub-optimal areas.
Furthermore, as shown in \cite{sonderby2016ladder}, the Stochastic RNN inherits the limitation of VAEs, which are difficult to optimize for deep hierarchies due to multiple layers of conditional stochastic layers.
The KL vanishing problem and the shallow structure can degrade the capacity of the Stochastic RNN, which is crucial to model highly variable temporal patterns.

\section{Proposed Method}
In this section, we first present the network architecture of our proposed H-VGRAE.
Then, we describe the offline and online stages in H-VGRAE, respectively.
\subsection{H-VGRAE}
\begin{figure}[htbp]
\centerline{\includegraphics[width=0.48\textwidth]{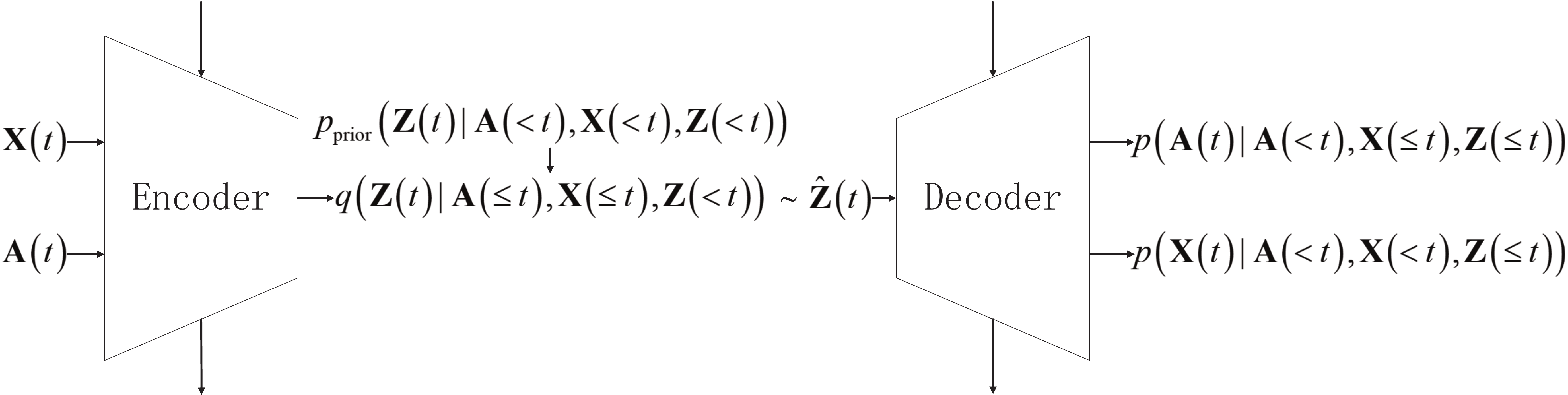}}
\caption{Architecture of H-VGRAE at timestamp $t$.}
\label{H-VGRAE_arc}
\end{figure}
H-VGRAE is in the general framework of stochastic RNN, which is composed of an encoder to encodes the input into low-dimensional latent representations and a decoder to reconstruct the input based on the latent representations, as shown in Fig.\ref{H-VGRAE_arc}.
The basic idea of H-VGRAE is the following.
First, it combines the GAE and RNN to capture ST patterns and content information from the dynamic of network topology and node attributes.
Second, the current ST features and historical state are encoded into low-dimensional random variables, which can model high variability of the data patterns.
Third, inspired by the utilization of multi-scale feature extractors in the speech processing literature \cite{chung2016hierarchical}, we design a hierarchical structure to extract multi-scale spatial-temporal features and encode them to multi-layer stochastic latent variables.
We then design a information sharing mechanism between random variables in the encoder and the decoder to alleviate the difficulties of optimizing multiple stochastic layers.
Forth, we employ the Normalizing Flow \cite{rezende2015variational} on the random variables to model non-Gaussian distributions.

The model of H-VGRAE is based on the likelihood maximizing like Stochastic RNN, but have an additional input of $\mathbf{X}(t)$.
For the $T$-length snapshots of $\mathbf{A}({\leq} T)$ and $\mathbf{X}({\leq} T)$, the likelihood with latent variable $\mathbf{Z}({\leq} T)$ is
\begin{equation}
  \begin{split}
  &\log p(\mathbf{A}({\leq} T),\mathbf{X}({\leq} T)) \\
  &\geq \mathbb{E}_{q(\mathbf{Z}({\leq} T))|\mathbf{A}({\leq} T),\mathbf{X}({\leq} T)} \Big[-\log q(\mathbf{Z}({\leq} T))|\mathbf{A}({\leq} T),\mathbf{X}({\leq} T)) \\
  &+ \log p(\mathbf{A}({\leq} T),\mathbf{X}({\leq} T),\mathbf{Z}({\leq} T))  \Big] \ \ \ \ \text{(ELBO)}
\end{split}
\label{elb}
\end{equation}
where,
\begin{equation}
  \begin{split}
  &q(\mathbf{Z}({\leq} T))|\mathbf{A}({\leq} T),\mathbf{X}({\leq} T))\\
  &=\prod\limits_{t = 1}^{T} {q(\mathbf{Z}(t)|\mathbf{A}({\leq} T),\mathbf{X}({\leq} T),\mathbf{Z}(\textless t))}\\
  &=\prod\limits_{t = 1}^{T} {q(\mathbf{Z}(t)|\mathbf{A}({\leq} t),\mathbf{X}({\leq} t),\mathbf{Z}(\textless t))} \ \ \ \ \text{(posterior)}
\end{split}
\end{equation}

\begin{equation}
  \begin{split}
  &p(\mathbf{A}({\leq} T),\mathbf{X}({\leq} T),\mathbf{Z}({\leq} T) )\\
  &=\prod\limits_{t = 1}^{T}  p_{\text{prior}}(\mathbf{Z}(t)|\mathbf{A}(\textless t),\mathbf{X}(\textless t),\mathbf{Z}(\textless t)) \ \ \ \ \text{(prior)}\\
  &\cdot p(\mathbf{X}(t)|\mathbf{A}(\textless t),\mathbf{X}(\textless t),\mathbf{Z}({\leq} t))\\
  &\cdot {p(\mathbf{A}(t)|\mathbf{A}(\textless t),\mathbf{X}(\textless t),\mathbf{Z}({\leq} t))} \ \ \ \ \text{(reconstruction)}
\end{split}
\label{joint}
\end{equation}

It is to be noticed that the ELBO in Eq. \ref{elb} is different from the loss function of VGRNN \cite{hajiramezanali2019variational}.
The loss function of VGRNN does not have the term of $p(\mathbf{X}(t)|\mathbf{A}(\textless t),\mathbf{X}(\textless t),\mathbf{Z}({\leq} t))$, which means it cannot guarantee to maximize the essential ELBO.

\begin{figure*}[htbp]
\centerline{\includegraphics[width=0.95\textwidth]{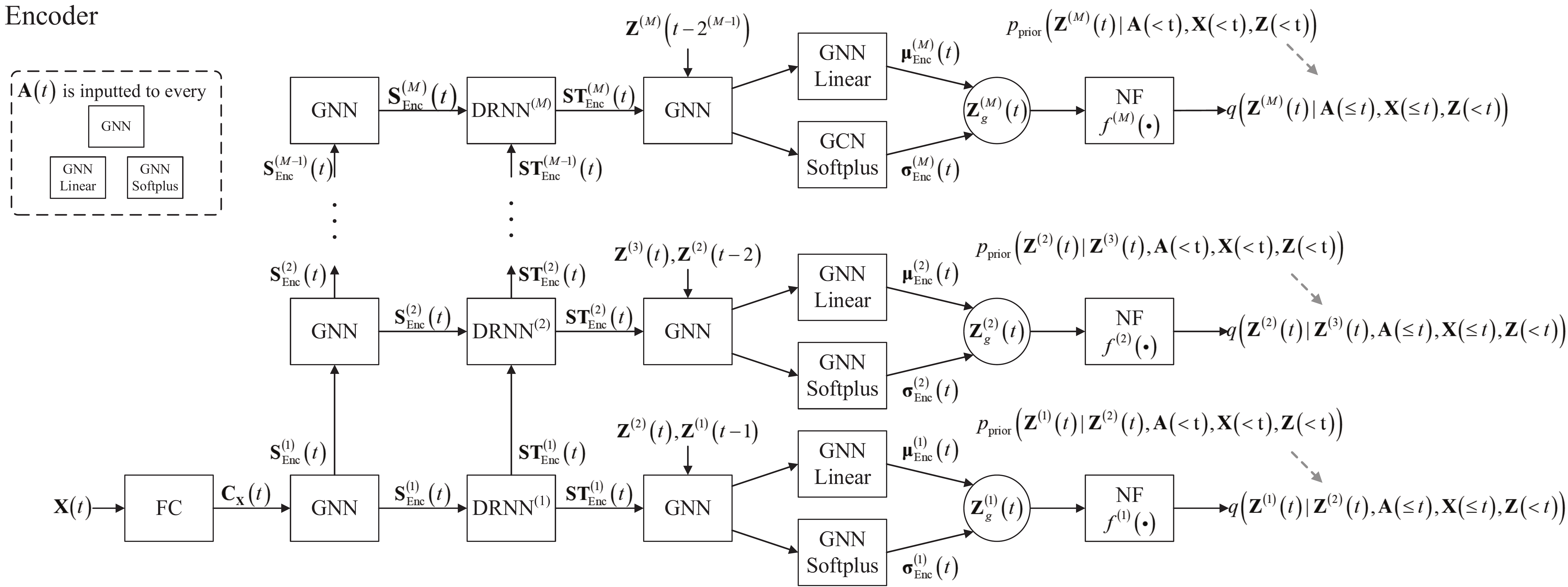}}
\caption{Structure of the encoder in H-VGRAE at timestamp $t$.}
\label{encoder}
\end{figure*}

\begin{figure*}[htbp]
\centerline{\includegraphics[width=0.95\textwidth]{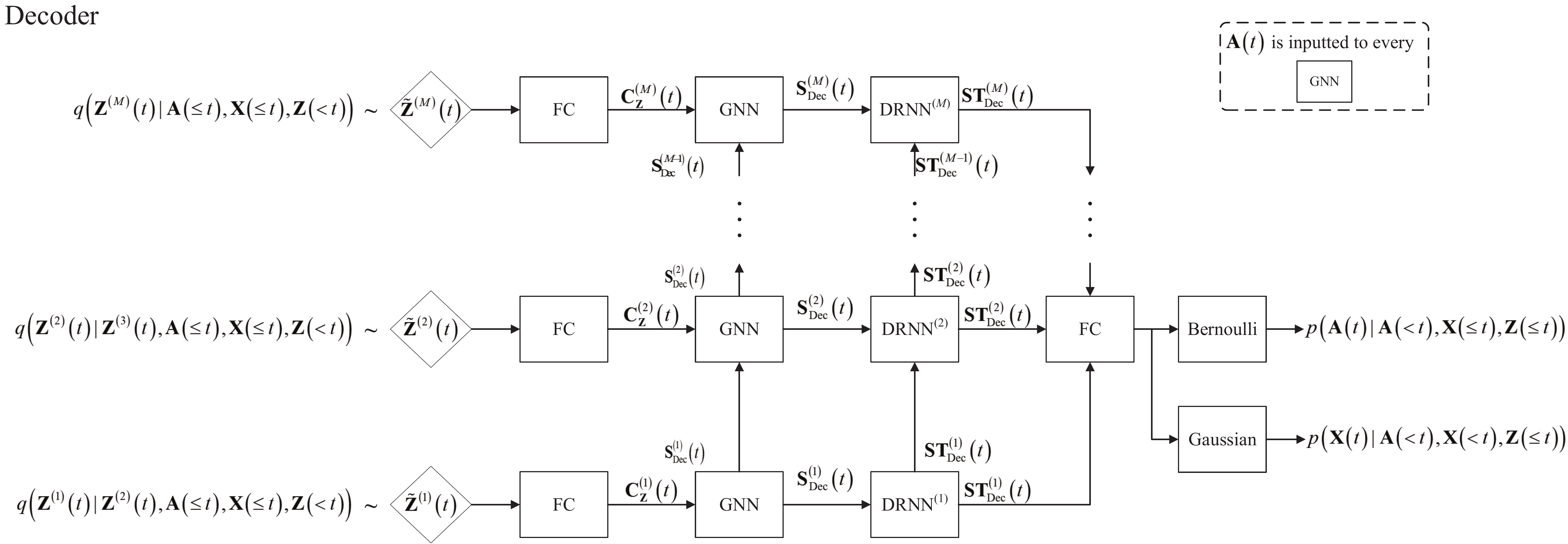}}
\caption{Structure of the decoder in H-VGRAE at timestamp $t$.}
\label{decoder}
\end{figure*}

The task is to maximize the ELBO by make estimations of the posterior, the prior and the reconstruction probabilities.
In H-VGRAE, the inference of the three probability distributions is implemented by an Encoder-Decoder framework.
A possible implementation of the inference network is composed of GAE and Stochastic RNN to extract spatial and temporal features respectively.
However, limited by the local approximation of GNN and the memory capability of RNN, the inference of the three distributions can only focus on local spatial features and short-term temporal features.
The limited reception field could cause inaccurate inference and make the anomaly detection sensitive to local and short-term changes.

To tackle this problem, we design a hierarchical spatial-temporal inference model to extract multi-scale ST features and encodes them to hierarchical stochastic latent variables.
The network architecture of the encoder and the decoder in H-VGRAE is shown in Fig. \ref{encoder} and \ref{decoder} respectively.
In the encoder, it stacks multiple GNN layers and dilated RNN (DRNN) layers \cite{drnn} to extract multi-scale ST and content features from $\mathbf{A}({\leq}t)$ and $\mathbf{X}({\leq}t)$, and then encodes current feature information and historical state into stochastic latent representation $\mathbf{Z}(t) = (\mathbf{Z}^{(1)}(t),\ldots,\mathbf{Z}^{(M)}(t))$, where $\mathbf{Z}^{(i)}(t) = (\mathbf{Z}^{(i)}_{1}(t),\ldots,\mathbf{Z}^{(i)}_{N}(t)) \in \mathbb{R}^{N \times d} (1{\leq}i{\leq}M)$, $d$ is the dimension of the latent variables.
In the decoder, it uses $\mathbf{Z}(t)$ to reconstruct the node attributes $\mathbf{X}({\leq}t)$ and the adjacency matrices $\mathbf{A}({\leq}t)$.
Similar with the encoder, the decoder first decodes the multi-scale content and ST features of $\mathbf{Z}(t)$, then reconstruct $\mathbf{A}({\leq}t)$ and $\mathbf{X}({\leq}t))$ by a Bernoulli MLP and a Gaussian MLP respectively.

\subsubsection{Encoder}
\

The structure of the encoder is shown in Fig.\ref{encoder}.
The encoder is composed of a content feature extraction block, a multi-scale ST feature extraction block and a hierarchical stochastic block.
The content feature extraction block $\varphi^{\mathbf{x}}(\cdot)$ is be implemented with MLP.
In the $M$-scale ST feature extraction block, there are $M$ stacked GNN layers, and each is followed by a DRNN layer.
Following each DRNN layer, the latent variable $\mathbf{Z}^{(i)}(t)$ of all nodes is inferred by a inference network \[\text{GNN}-(\text{GNN Linear}, \text{GNN Softplus})\] where the activation functions of GNN, GNN Linear and GNN Softplus are ReLU, none and Softplus.

The procedure of the encoder at time $t$ is as follows.

\textbf{Step 1}: The node attributes $\mathbf{X}(t)$ are inputted to $\varphi^{\mathbf{X}}(\cdot)$ to extract content features $\mathbf{C}_{\mathbf{X}}(t) = \varphi^{\mathbf{X}}(\mathbf{X}(t))$.
$\varphi^{\mathbf{X}}(\cdot)$ are deep neural networks which operate on each node independently, and are crucial for learning complex content features.

\textbf{Step 2}: $\mathbf{C}_{\mathbf{X}}(t)$ are inputted into a sequence of GNNs ($M$ in total), and $i$-th scale of spatial features is
\[{\mathbf{S}}_{{\text{Enc}}}^{(i)}(t) = \text{GNN}({\mathbf{S}}_{{\text{Enc}}}^{(i-1)}(t),\mathbf{A}(t))\]
where ${\mathbf{S}}_{{\text{Enc}}}^{(0)}(t) = \mathbf{C}_{\mathbf{X}}(t)$.

The GNN can be implemented with GCN \cite{GCN}, whose calculation process in the $i$-th layer can be formulated as follows:
\begin{equation}
  \label{GNN}
  \mathbf{S}_{\text{Enc}}^{(i)}(t) = \sigma (\mathbf{\hat{A}}(t){\mathbf{S}}_{{\text{Enc}}}^{(i-1)}(t)\mathbf{W}^{(i)}),
\end{equation}
where $\sigma(\cdot)$ is a activation function with non-linearity.
In our model, we choose the $\text{ReLU}(x) = \max (0,x)$ to be the activation function.
$\mathbf{\hat{A}}(t) = \mathbf{\tilde{D}}^{-\frac{1}{2}}(t)\mathbf{\tilde{A}}(t)\mathbf{\tilde{D}}^{-\frac{1}{2}}(t)$ is the regularized adjacency matrix with self loops, where $\mathbf{\tilde{A}}(t) = \mathbf{A}(t) + \mathbf{I}_{N \times N}$ is the adjacency matrix with self loops and $\mathbf{\tilde{D}}$ is the diagonal degree matrix, where $\tilde{D}_{ii}(t) = \sum\nolimits_{j} \tilde{A}_{ij}(t)$ denotes the degree of node $i$ and $\tilde{D}_{ik}(t) = 0$ for $k \not= i$.
A single layer of GNN can represent each node by fusing the data of this node and its $1$-hop neighbors.
As proved in \cite{GCN}, stacking $i$ layers of GNN is equivalent to represent each node by considering its $1$-hop to $i$-hop neighbors' data, which is corresponding with the $i$-th spatial scale in our model.


\textbf{Step 3}: Each $\mathbf{S}_{\text{Enc}}^{(i)}(t)$ and $\mathbf{ST}_{\text{Enc}}^{(i-1)}(t)$ are concatenated to $\mathbf{S'}_{\text{Enc}}^{(i)}(t) = [\mathbf{S}_{\text{Enc}}^{(i)}(t) || \mathbf{ST}_{\text{Enc}}^{(i-1)}(t)]$ for $(i=2,\dots,M)$ and $\mathbf{S'}_{\text{Enc}}^{(i)}(t) = \mathbf{S}_{\text{Enc}}^{(i)}(t)$ for $(i=1)$. $\mathbf{S'}_{\text{Enc}}^{(i)}(t)$ is inputted into $\text{DRNN}^{(i)}(\cdot)$ to capture the $i$-th scale of ST features $\mathbf{ST}_{\text{Enc}}^{(i)}(t) = \text{DRNN}^{(i)}(\mathbf{S'}_{\text{Enc}}^{(i)}(t))$.

The correspondence of spatial scales and temporal scales is based on the nature of the dynamic networks, that the information propagation process is controlled by the distance between nodes.
Supposing that the messages transmitted by some node include both its own messages and its neighbors' messages, and the time consumed during transmission is only proportional to the hops between two nodes.
The more distant nodes which are considered in upper GNNs should have less transmission frequencies, which means those messages should correspond to the long-term trend.

\begin{figure}[t]
\centerline{\includegraphics[width=0.48\textwidth]{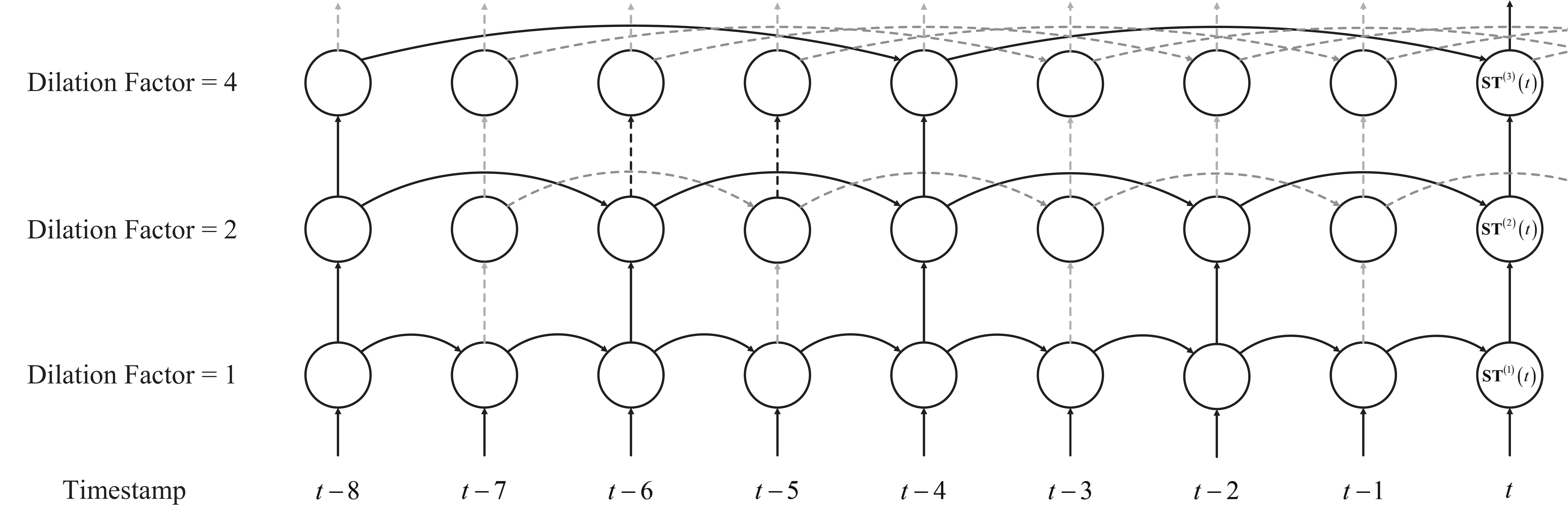}}
\caption{An example of a three-layer DRNN with dilation factor 1, 2, and 4.}
\label{HDRNN}
\end{figure}

To model these different frequencies, $\text{DRNN}^{(i)}(\cdot)$ \cite{drnn} are used in the H-VGRAE.
The structure of a three-layer $DRNN$ with exponentially increased dilation factors $1$, $2$, and $4$ is shown in Fig.\ref{HDRNN}.
By stacking multiple DRNN layers, the reception fields are extended in different scales.
The hidden units at different timestamps are connected based on the layer level, in which the hidden unit $\mathbf{h}^{(i)}(t-2^{i-1})$ is connected with $\mathbf{h}^{(i)}(t)$.
By such skip connections, the stacked DRNNs can extract multi-scale temporal features and alleviate the vanishing  gradient problem of standard RNN.

$\text{DRNN}^{(i)}(\cdot)$ can be implemented with two popular variants of RNNs, which are Long Short-Term Memory (LSTM) \cite{hochreiter1997long} and Gated Recurrent Units (GRU) \cite{chung2014empirical}.
In general, the performance of GRU is as good as LSTM and GRU has fewer parameters and simpler structure.
In the case of using GRU, the $\text{DRNN}^{(i)}(\cdot)$ for node $k$ is formulated as follows:
\begin{equation*}
\begin{split}
  &\mathbf{r}_{k}^{(i)}(t) = \sigma (\mathbf{W}^{(i)}_{r_k} [\mathbf{h}^{(i)}_{k}(t-2^{i-1}),\mathbf{S'}_{\text{Enc},k}^{(i)}(t)]) \\
  &\mathbf{u}_{k}^{(i)}(t) = \sigma (\mathbf{W}^{(i)}_{u_k} [\mathbf{h}^{(i)}_{k}(t-2^{i-1}),\mathbf{S'}_{\text{Enc},k}^{(i)}(t)]) \\
  &\mathbf{\tilde{h}}_{k}^{(i)}(t) = tanh(\mathbf{W}^{(i)}_{\tilde{h}_k} [\mathbf{r}_{k}^{(i)}(t) \odot \mathbf{h}^{(i)}_{k}(t-2^{i-1}),\mathbf{S'}_{\text{Enc},k}^{(i)}(t)]) \\
  &\mathbf{h}_{k}^{(i)}(t) = (1 - \mathbf{u}_{k}^{(i)}(t)) \odot \mathbf{h}^{(i)}_{k}(t-2^{i-1}) + \mathbf{u}_{k}^{(i)}(t) \odot \mathbf{S'}_{\text{Enc},k}^{(i)}(t)\\
  &\mathbf{ST}_{\text{Enc},k}^{(i)}(t) = \mathbf{h}_{k}^{(i)}(t)
\end{split}
\end{equation*}


\textbf{Step 4}: $\mathbf{G}^{(M)}(t)=[\mathbf{ST}_{\text{Enc}}^{(M)}(t)||\mathbf{Z}^{M}(t-2^{M-1})]$ is inputted into a inference network $\text{GNN}-(\text{GNN Linear}, \text{GNN Softplus})$ to infer the Gaussian posterior distribution in the largest scale $\mathbf{Z}^{(M)}_{g}(t)$,
\begin{equation}
\begin{split}
&q(\mathbf{Z}^{(M)}_{g}(t)|A({\leq}t),X({\leq}t),\mathbf{Z}(t)) \\
&= \prod_{k=1}^{N}q(\mathbf{Z}^{(M)}_{g,k}(t)|A({\leq}t),X({\leq}t),\mathbf{Z}(t)) \\
&= \prod_{k=1}^{N} \mathcal{N}(\boldsymbol{\mu}_{\text{Enc},k}^{(M)}(t),\boldsymbol{\sigma}_{\text{Enc},k}^{2,(M)}(t)\mathbf{I}) \\
\end{split}
\label{Zenc}
\end{equation}
where
\begin{equation}
\begin{split}
&\boldsymbol{\mu}_{\text{Enc},k}^{(M)}(t) = \text{GNN}(\text{GNN Linear}(\mathbf{G}^{(M)}(t))\\
&\boldsymbol{\sigma}_{\text{Enc},k}^{2,(M)}(t) = \text{GNN}(\text{GNN Softplus}(\mathbf{G}^{(M)}(t))
\end{split}
\label{Zenc}
\end{equation}
Then, for each $i \textless M$, in addition to $\mathbf{ST}_{\text{Enc}}^{(i)}(t),\mathbf{Z}^{(i)}(t-2^{M-1})$, the inference of $\mathbf{Z}^{(i)}_{g}(t)$ can also leverage the information of the $(i+1)$-th latent variable $\mathbf{Z}^{(i+1)}(t)$.
The up-down structure can pass the high-level information to the low-level feature extraction, to assist the inference local and short-term latent variable with global and long-term information.
Then the following inference of $\mathbf{Z}^{(i)}_{g}(t)$ is the same as Eq.\ref{Zenc}.




\textbf{Step 5}: Gaussian $\mathbf{Z}^{(i)}_{g}(t)$ is extended to non-Gaussian $\mathbf{Z}^{(i)}(t)$, by inputting $\mathbf{Z}^{(i)}_{g}(t)$ to a chain of normalizing flow (NF) functions $f^{(i)}(\cdot)$.

According to the normalizing flow technique \cite{rezende2015variational}, if $\mathbf{Z}^{(i)}_{k}(t) = f(\mathbf{Z}^{(i)}_{g,k}(t))$ for each node $k$ and $f(\cdot)$ is invertible and smooth, the distribution of $\mathbf{Z}^{(i)}_{k}(t)$ is then
\begin{equation}
  \begin{split}
  &q(\mathbf{Z}^{(i)}_{k}(t)|A({\leq}t),X({\leq}t),\mathbf{Z}(t)) \\
  &= q(\mathbf{Z}^{(i)}_{g,k}(t)|A({\leq}t),X({\leq}t),\mathbf{Z}(t)) \left|{\text{det}\frac{\partial f^{-1}}{\partial \mathbf{Z}^{(i)}_{k}(t)}} \right| \\
  &= q(\mathbf{Z}^{(i)}_{g,k}(t)|A({\leq}t),X({\leq}t),\mathbf{Z}(t)) \left|{\text{det}\frac{\partial f}{\partial \mathbf{Z}^{(i)}_{g,k}(t)}} \right|^{-1}
\end{split}
\end{equation}
There are various choices of $f(\cdot)$.
A popular and simple one is planar NF, in which
\begin{equation}
  f(\mathbf{Z}^{(i)}_{g,k}(t)) = \mathbf{Z}^{(i)}_{g,k}(t) + \mathbf{W}_{f,1}tanh(\mathbf{W}_{f,2}\mathbf{Z}^{(i)}_{g,k}(t) + \mathbf{b}_{f,2})
\end{equation}
To make the non-Gaussian $\mathbf{Z}^{(i)}(t)$ more variable, the planar NF can be treated as a layer and be stacked for $L$ times $f^{(i)}(\mathbf{Z}^{(i)}_{g,k}(t))=f^{L}(f^{L-1}(\ldots f^{1}(\mathbf{Z}^{(i)}_{g,k}(t))))$.
This extension enhances the representation ability of the latent representation.

The up-down structure leads the inference of the posterior be factorized as
\begin{equation}
  \begin{split}
  &q(\mathbf{Z}(t)|\mathbf{A}({\leq}t),\mathbf{X}({\leq}t),\mathbf{Z}(\textless t))\\
  &= q(\mathbf{Z}^{(M)}(t)|\mathbf{A}({\leq}t),\mathbf{X}({\leq}t),\mathbf{Z}(\textless t)) \\
  &\cdot \prod_{i=1}^{M-1} q(\mathbf{Z}^{(i)}(t)|\mathbf{Z}^{(i+1)}(t),\mathbf{A}({\leq}t),\mathbf{X}({\leq}t),\mathbf{Z}(\textless t))\\
  &= \prod_{k=1}^{N} q(\mathbf{Z}_{k}^{(M)}(t)|\mathbf{A}({\leq}t),\mathbf{X}({\leq}t),\mathbf{Z}(\textless t))\\
  &\cdot \prod_{i=1}^{M-1}\prod_{k=1}^{N} q(\mathbf{Z}^{(i)}_{k}(t)|\mathbf{Z}^{(i+1)}(t),\mathbf{A}({\leq}t),\mathbf{X}({\leq}t),\mathbf{Z}(\textless t))
\end{split}
\end{equation}

\subsubsection{Decoder}
\

The structure of the decoder is nearly symmetric of that of the encoder, as shown in Fig.\ref{decoder}.
The input $\tilde{\mathbf{Z}}^{(i)}(t)$ to the decoder is sampled from $q(\mathbf{Z}^{(i)}(t)|\mathbf{Z}^{(i+1)}(t),\mathbf{A}({\leq}t),\mathbf{X}({\leq}t),\mathbf{Z}(\textless t))$ for $i \textless M$ and $q(\mathbf{Z}^{(i)}(t)|\mathbf{A}({\leq}t),\mathbf{X}({\leq}t),\mathbf{Z}(\textless t))$ for $i=M$.
The decoder is similar with that of the encoder is also composed of a content feature extraction block, multi-scale spatial-temporal blocks and density estimators on the probability of $\mathbf{A}(t)$ and $\mathbf{X}(t)$.
Considering the data structure, we choose Bernoulli MLP to reconstruct $\mathbf{A}(t)$ and Gaussian MLP to reconstruct $\mathbf{X}(t)$.
\begin{figure}[t]
\centerline{\includegraphics[width=0.48\textwidth]{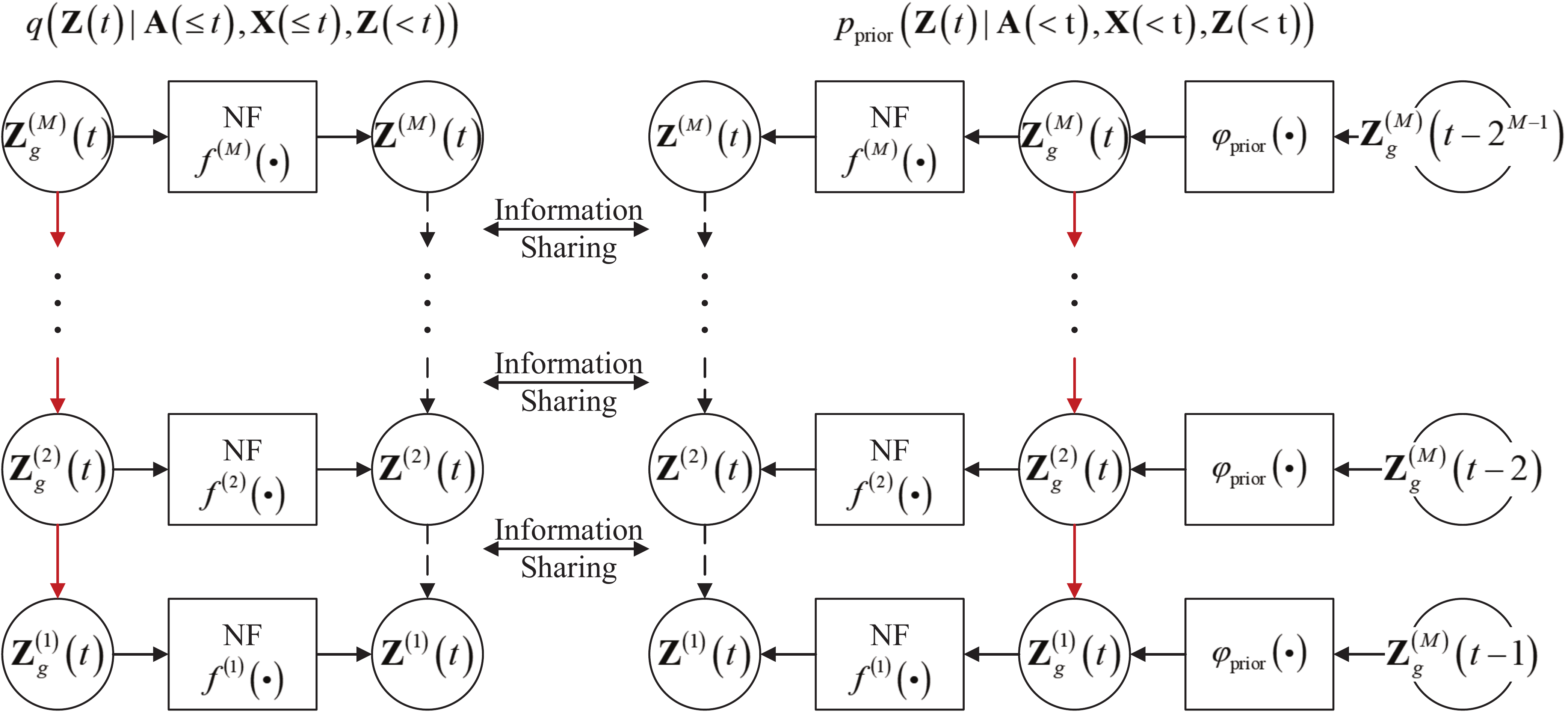}}
\caption{The information sharing mechanism between the inferred posterior and the prior in H-VGRAE.}
\label{information_sharing}
\end{figure}

The procedure of the decoder at time t is as follows.

\textbf{Step 1}: Sample $\tilde{\mathbf{Z}}^{(i)}(t)$ from the inferred posterior.

It is difficult to sample $\mathbf{Z}^{(i)}(t)$ directly, due to its complex and implicit distribution.
To tackle this problem, we can first sample a value $\tilde{\mathbf{Z}}^{(i)}_{g,k}(t)$ of $\mathbf{Z}^{(i)}_{g,k}(t)\sim \mathcal{N}(\boldsymbol{\mu}_{\text{Enc},k}^{(i)}(t),\boldsymbol{\sigma}_{\text{Enc},k}^{2,(i)}(t)\mathbf{I})$ easily, using the reparametrization trick.
Let $\mathbf{Z}^{(i)}_{g,k}(t) = \boldsymbol{\mu}_{\text{Enc},k}^{i}(t) + \boldsymbol{\varepsilon} \odot \boldsymbol{\sigma}_{\text{Enc},k}^{(i)}(t)$, where $\boldsymbol{\varepsilon} \sim \mathcal{N}(\mathbf{0},\mathbf{I})$.
Then, the sampled value $\tilde{\mathbf{Z}}^{(i)}_{g,k}(t)$ is inputted the planar NF function to obtain $\tilde{\mathbf{Z}}^{(i)}_{k}(t) = f^{(i)}(\tilde{\mathbf{Z}}^{(i)}_{g,k}(t))$.

\textbf{Step 2}: Input $\tilde{\mathbf{Z}}^{(i)}(t)$ to the content feature block (FC layers) $\varphi^{\mathbf{Z}}(\cdot)$ to decode the latent content features $\mathbf{C}_{\mathbf{Z}}^{(i)}(t) = \varphi^{\mathbf{Z}}(\tilde{\mathbf{Z}}^{(i)}(t)) $ at each node.

\textbf{Step 3}: Input $\mathbf{C}_{\mathbf{Z}}^{(i)}(t)$ to the $i$-the layer of GNN in the decoder to decode the latent spatial features $\mathbf{S}_{\text{Dec}}^{(i)}(t)$.
The computation process is the similar with the decoder in Eq.\ref{GNN}.


\textbf{Step 4}: Each $\mathbf{S}_{\text{Dec}}^{(i)}(t)$ and $\mathbf{ST}_{\text{Dec}}^{(i-1)}(t)$ are concatenated to $\mathbf{S'}_{\text{Dec}}^{(i)}(t) (\mathbf{S'}_{\text{Dec}}^{(1)}(t) = \mathbf{S}_{\text{Dec}}^{(i)}(t))$ to the $\text{DRNN}^{(i)}(\cdot)$ in the decoder to decode the latent ST features $\mathbf{ST}_{\text{Dec}}^{(i)}(t)$.


\textbf{Step 5}: Input the decoded ST features to a FC layer to fuse them, and use the fused features $\mathbf{F}(t)$ to reconstruct adjacency matrix $\mathbf{A}(t)$.
Considering the data structure, we use a Bernoulli MLP to obtain $p(\mathbf{A}(t)|\mathbf{A}(\textless t),\mathbf{X}({\leq} t),\mathbf{Z}({\leq} t))$ and a Gaussian MLP to obtain $p(\mathbf{X}(t)|\mathbf{A}(\textless t),\mathbf{X}({\leq} t),\mathbf{Z}({\leq} t))$.

The Bernoulli MLP utilized is in the simple form:
\begin{equation*}
  \begin{split}
  &\log p(\mathbf{A}(t)|\mathbf{A}(\textless t),\mathbf{X}({\leq} t),\mathbf{Z}({\leq} t))\\
  &= \sum_{i=1}^{N}\sum_{j=1}^{N}
  A_{ij}(t)\log y_{ij}(t) + (1-A_{ij}(t))\log (1-y_{ij}(t))
\end{split}
\end{equation*}
where
\begin{equation*}
  \mathbf{Y}_{k}(t) = \text{Sigmoid}(\mathbf{W}_{2}\tanh(\mathbf{W}_{1}\mathbf{F}_{k}(t)+\mathbf{b}_{1})+\mathbf{b}_{2})
\end{equation*}
and the Gaussian MLP is:
\begin{equation*}
  \begin{split}
  &\log p(\mathbf{X}(t)|\mathbf{A}(\textless t),\mathbf{X}({\leq} t),\mathbf{Z}({\leq} t))\\
  &= \sum_{k=1}^{N}
  \log p(\mathbf{x}_{k}(t)|\mathbf{A}(\textless t),\mathbf{X}({\leq} t),\mathbf{Z}({\leq} t))\\
  &= \sum_{k=1}^{N} \log \mathcal{N}(\boldsymbol{\mu}_{\mathbf{x}_{k}}(t),\boldsymbol{\sigma}_{\mathbf{x}_{k}}^2(t)\mathbf{I})
\end{split}
\end{equation*}
where
\begin{equation*}
  \begin{split}
  &\boldsymbol{\mu}_{\mathbf{x}_{k}}(t) = \mathbf{W}_{4}\text{tanh}(\mathbf{W}_{3} \mathbf{F}_{k}(t) + \mathbf{b}_{3}) + \mathbf{b}_{4}\\
  &\boldsymbol{\sigma}_{\mathbf{x}_{k}}^2(t) = \mathbf{W}_{5}\text{tanh}(\mathbf{W}_{3} \mathbf{F}_{k}(t) + \mathbf{b}_{3}) + \mathbf{b}_{5}\\
\end{split}
\end{equation*}
\subsubsection{Prior}
\

As shown in Eq.\ref{SRNN}, the decoder need to generate a prior distribution of the latent variable.
In static scenarios, the prior is commonly set to be $\mathcal{N}(\mathbf{0},\mathbf{I})$ \cite{VAE}.
However, as shown in Stochastic RNNs, $\mathcal{N}(\mathbf{0},\mathbf{I})$ would be too simple to capture temporal information in the scenario of dynamic networks \cite{chung2015recurrent}.
Moreover, utilizing complex decoders like RNNs may lead to the KL term in Eq.\ref{SRNN} vanishing \cite{bowman2016generating}.
This means the latent variable and the input are independent.
The encoder $q(\mathbf{Z}(t)|\mathbf{A}({\leq} t),\mathbf{X}({\leq} t),\mathbf{Z}(\textless t)) = q(\mathbf{Z}(t))$ and the decoder $p(\mathbf{A}(t)|\mathbf{A}(\textless t),\mathbf{X}({\leq} t),\mathbf{Z}({\leq} t)) = p(\mathbf{A}(t)|\mathbf{X}(t))$ would just fit the dataset.
This makes the training ends to a two-step learning procedure and leads to sub-optimal model.

To avoid such problem, we let the prior $p_{\text{prior}}(\mathbf{Z}(t)|\mathbf{A}(\textless t),\mathbf{X}(\textless t),\mathbf{Z}(\textless t))$ come from the prediction based on previous timestamps.
Similarly, we first generate the Gaussian form,
\[p_{prior}(\mathbf{Z}^{(i)}_{g,k}(t)|\mathbf{A}(\textless t),\mathbf{X}(\textless t),\mathbf{Z}(\textless t)) =  \mathcal{N}(\boldsymbol{\mu}_{\text{prior},k}^{(i)}(t),\boldsymbol{\sigma}_{\text{prior},k}^{2,(i)}(t)\mathbf{I}),\]
where $\{\boldsymbol{\mu}_{\text{prior},k}^{(i)}(t),\boldsymbol{\sigma}_{\text{prior},k}^{2,(i)}(t)\} = \varphi_{\text{Prior}}(\mathbf{Z}^{(i)}_{g}(t-2^{i-1}))$, $\varphi_{\text{Prior}}(\cdot)$ can also be implemented with DRNN.
The Gaussian prior $\mathbf{Z}^{(i)}_{\text{prior},g}(t)$ is also extended to non-Gaussian prior $\mathbf{Z}^{(i)}(t)$ by using the same planar NF in the encoder.
Benefit from the flexible settings of the posterior and prior, the KL vanishing problem would be alleviated.

Similar with the posterior, the prior is also be factorized as
\begin{equation}
  \begin{split}
   &p_{\text{prior}}(\mathbf{Z}(t)|\mathbf{A}(\textless t),\mathbf{X}(\textless t),\mathbf{Z}(\textless t))\\
   &= \prod_{k=1}^{N} p_{\text{prior}}(\mathbf{Z}_{k}^{(M)}(t)|\mathbf{A}(\textless t),\mathbf{X}(\textless t),\mathbf{Z}(\textless t))\\
    &\cdot \prod_{i=1}^{M-1} p_{\text{prior}}(\mathbf{Z}^{(i)}_{k}(t)|\mathbf{Z}^{(i+1)}(t),\mathbf{A}(\textless t),\mathbf{X}(\textless t),\mathbf{Z}(\textless t))
\end{split}
\end{equation}

In the view of unsupervised representation learning, the predictive prior is an auxiliary objective for the latent variables.
Furthermore, due to the different scales of the priors, the prediction process is also hierarchical.
Ideally, with the help of the predictive prior, the encoder can infer not only the best latent representation to reconstruct current input, but also the best to forecast future snapshots.

Unfortunately, as illustrated in \cite{sonderby2016ladder}, highly flexible models with deep hierarchies would be difficult to optimize due to multiple conditional stochastic layers.
To tackle such problem in H-VGRAE, we aim to find more relations between the encoder and the decoder to help the optimization.
Following the information flow direction, $\mathbf{Z}^{(i)}(t)$ should be conditioned on $\mathbf{Z}^{(i+1)}(t)$ $(i\textless M)$.
As shown in Fig.\ref{information_sharing}, both the encoder and the prior should infer the distributions of $\mathbf{Z}^{(i)}(t)|\mathbf{Z}^{(i+1)}(t)$, but obtain $q(\mathbf{Z}^{(i)}(t)|\mathbf{Z}^{(i+1)}(t),\mathbf{A}({\leq}t),\mathbf{X}({\leq}t),\mathbf{Z}(\textless t))$ and $p_{\text{prior}}(\mathbf{Z}^{(i)}(t)|\mathbf{Z}^{(i+1)}(t),\mathbf{A}(\textless t),\mathbf{X}(\textless t),\mathbf{Z}(\textless t))$ respectively.
If the encoder and the decoder could share information about the two distributions, the training can be easier to converge.

However, the posterior and prior cannot be explicitly modeled, due to non-Gaussian property by the NF.
To achieve the goal of information sharing, we have to trace back to the Gaussian form.
Considering the NF, the posterior can be rewritten as
\begin{equation}
  \begin{split}
  &q(\mathbf{Z}^{(i)}_{k}(t)|\mathbf{Z}^{(i+1)}_{k}(t),\mathbf{A}({\leq}t),\mathbf{X}({\leq}t),\mathbf{Z}(\textless t)) \\
  &= q(f^{(i)}(\mathbf{Z}^{(i)}_{g,k}(t))|f^{(i+1)}(\mathbf{Z}^{(i+1)}_{g,k}(t)),\mathbf{A}({\leq}t),\mathbf{X}({\leq}t),\mathbf{Z}(\textless t))\\
  &= q(f^{(i)}(\mathbf{Z}^{(i)}_{g,k}(t))|\mathbf{Z}^{(i+1)}_{g,k}(t),\mathbf{A}({\leq}t),\mathbf{X}({\leq}t),\mathbf{Z}(\textless t))\\
  &= q(\mathbf{Z}^{(i)}_{g,k}(t)|\mathbf{Z}^{(i+1)}_{g,k}(t),\mathbf{A}({\leq}t),\mathbf{X}({\leq}t),\mathbf{Z}(\textless t))\left|{\text{det}\frac{\partial f^{i}}{\partial \mathbf{Z}^{(i)}_{g,k}(t)}} \right|^{-1}
\end{split}
\end{equation}
Similarly,
\begin{equation}
  \begin{split}
  &p_{\text{prior}}(\mathbf{Z}^{(i)}_{k}(t)|\mathbf{Z}^{(i+1)}_{k}(t),\mathbf{A}(\textless t),\mathbf{X}(\textless t),\mathbf{Z}(\textless t)) \\
  &=p_{\text{prior}}(\mathbf{Z}^{(i)}_{g,k}(t)|\mathbf{Z}^{(i+1)}_{g,k}(t),\mathbf{A}(\textless t),\mathbf{X}(\textless t),\mathbf{Z}(\textless t))\left|{\text{det}\frac{\partial f^{i}}{\partial \mathbf{Z}^{(i)}_{g,k}(t)}} \right|^{-1}
  \end{split}
\end{equation}
For the inferred Gaussian distributions \[\mathcal{N}(\boldsymbol{\mu}_{\text{Enc},k}^{(i)}(t),\boldsymbol{\sigma}_{\text{Enc},k}^{2,(i)}(t)\mathbf{I})\] and \[\mathcal{N}(\boldsymbol{\mu}_{\text{prior},k}^{(i)}(t),\boldsymbol{\sigma}_{\text{prior},k}^{2,(i)}(t)\mathbf{I}) \]
We then can employ information sharing mechanism due to their explicit distribution representations.

Inspired by \cite{sonderby2016ladder}, we use the precision-weighted mechanism to introduce prior information into the encoder.
The updated mean and variance of  $q(\mathbf{Z}^{(i)}_{g,k}(t)|\mathbf{Z}^{(i+1)}_{g,k}(t),\mathbf{A}({\leq}t),\mathbf{X}({\leq}t),\mathbf{Z}(\textless t))$ is then,
\begin{subequations}
\begin{align}
  \bar{\boldsymbol{\sigma}}_{\text{Enc},k}^{(i)}(t) &= \frac{1}{\boldsymbol{\sigma}_{\text{Enc},k}^{-2,(i)}(t) + \boldsymbol{\sigma}_{\text{prior},k}^{-2,(i)}(t)} \\
  \bar{\boldsymbol{\mu}}_{\text{Enc},k}^{(i)}(t) &= \frac{\boldsymbol{\mu}_{\text{Enc},k}^{(i)}(t)\boldsymbol{\sigma}_{\text{Enc},k}^{-2,(i)}(t) + \boldsymbol{\mu}_{\text{prior},k}^{(i)}(t)\boldsymbol{\sigma}_{\text{prior},k}^{-2,(i)}(t)}{\boldsymbol{\sigma}_{\text{Enc},k}^{-2,(i)}(t) + \boldsymbol{\sigma}_{\text{prior},k}^{-2,(i)}(t)},
\end{align}
\end{subequations}
where $k$ means the $k$-th node.
In such a way, the optimization process would be easier.


\subsection{Offline Model Training}
\

The encoder and the decoder of H-VGRAE are simultaneously trained to tune the model parameters on the training set.
The ELBO is as shown in Eq. \ref{elb}, which can be computed by Monte Carlo integration
\begin{equation}
  \label{ELBO}
  \begin{split}
  &\mathcal{L}(T) \approx\\
  &\frac{1}{L} \sum\limits_{t=1}^{T}  \sum\limits_{l=1}^{L} \log p(\mathbf{A}(t)|\mathbf{A}(\textless t),\mathbf{X}({\leq} t),\tilde{\mathbf{Z}}_{l}({\leq} t)) \\
  &+\log p(\mathbf{X}(t)|\mathbf{A}(\textless t),\mathbf{X}(\textless  t),\tilde{\mathbf{Z}}_{l}({\leq} t) )  \\
  & + \log p_{\text{prior}}(\tilde{\mathbf{Z}}_{l}|\mathbf{A}(\textless t),\mathbf{X}(\textless  t),\tilde{\mathbf{Z}}_{l}(\textless  t))\\
  &- \log q(\tilde{\mathbf{Z}}_{l}(t)|\mathbf{A}({\leq} t),\mathbf{X}({\leq} t),\tilde{\mathbf{Z}}_{l}(\textless t)
  \end{split}
\end{equation}
where $\tilde{\mathbf{Z}}_{l}(t)$ is sampled from  $q(\mathbf{Z}(t)|\mathbf{A}({\leq} t),\mathbf{X}({\leq} t),\mathbf{Z}(\textless t))$.


After training H-VGRAE, the anomaly thresholds of $p(\mathbf{A}_{ij}(t)|\mathbf{A}(\textless t),\mathbf{X}({\leq} t),\mathbf{Z}({\leq} t))$  and $p(\mathbf{x}_{k}(t)|\mathbf{A}(\textless t),\mathbf{X}({\leq} t),\mathbf{Z}({\leq} t))$ should be determined based on the training set.
Let $\alpha_{\mathbf{A}}$ and $\alpha_{\mathbf{X}}$ denote the anomaly threshold of $p(\mathbf{A}_{ij}(t)|\mathbf{A}(\textless t),\mathbf{X}({\leq} t),\mathbf{Z}({\leq} t))$ and $p(\mathbf{x}_{k}(t)|\mathbf{A}(\textless t),\mathbf{X}({\leq} t),\mathbf{Z}({\leq} t))$ respectively.
Inspired by \cite{su2019robust}, we set the anomaly threshold following the principle of Extreme Value Theory (EVT) \cite{siffer2017anomaly} to achieve automatic threshold selection.

\subsection{Online Detection}
For an observation of the dynamic network $\mathcal{G}(t')$ at timestamp $t'$, 
%
by inputting $\mathbf{A}(t')$ and $\mathbf{X}(t')$ to the trained H-VGRAE, the reconstruction probabilities of edges and node attributes can be obtained.
If $\log p(A_{ij}(t')|\mathbf{A}(\textless t'),\mathbf{X}(\textless t'),\mathbf{Z}({\leq} t'))  \textless  \alpha_{\mathbf{A}}$, the edge for node $i$ to node $j$ is determined as anomalous; otherwise, it is normal.

For detecting anomalous nodes, there are two reasons causing a node to be anomalous, which are anomalous attributes and the edges from or to these nodes.
The anomalous nodes caused by the anomalous edges can be detected as above.
The anomalous nodes caused by anomalous attributes can be detected by comparing $p(\mathbf{x}_{k}(t)|\mathbf{A}(\textless t),\mathbf{X}({\leq} t),\mathbf{Z}({\leq} t))$ with $\alpha_{\mathbf{X}}$.

\subsection{Anomaly Interpretation}
The goal of anomaly interpretation is to interpret what caused an edge or a node to be an anomaly in terms of a comparable and illustrative metirc.
Methods in \cite{yu2018netwalk} and \cite{zheng2019addgraph} give illustration of the anomalous edges with heuristically defined anomaly scores.
However, these anomaly scores are not normalized, which makes the interpretation and comparison difficult.

In our model, the anomaly score is the reconstruction probability, which is in the range of $([0,1])$.
Because each edge and node are conditional independent in the reconstruction, we can rank all the anomaly scores of the edges and nodes, and locate them with low reconstruction probabilities.

Furthermore, to evaluate the anomalous levels of different dynamic networks, we can calculate the anomaly score in different granularity.
In the coarse granularity, we can use the proportion of the anomalous edges $\gamma_{e}(t') = |E_{a}(t')|/|E(t')|$ and the anomalous nodes $\gamma_{n}(t') = |V_{a}(t')|/|V(t')|$ as the anomaly coefficient, where $|E_{a}(t')|$ and $|E(t')|$ are the number of detected anomalous edges and total edges, $|V_{a}(t')|$ and $|V(t')|$ are the number of detected anomalous nodes and total nodes.
In the fined granularity, we can calculate the difference between $p(\mathbf{A}(t')|\mathbf{A}(\textless t'),\mathbf{X}(\textless t'),\mathbf{Z}({\leq} t'))$ and the distribution $p(\mathbf{A}^{thres}(t'))$ with each element $p(A^{thres}_{ij}(t')) = \alpha$.
The anomaly coefficient of edges can be calculated by using KL divergence \[\eta_{e}(t') = KL(p(\mathbf{A}(t')|\mathbf{A}(\textless t'),\mathbf{X}(\textless t'),\mathbf{Z}({\leq} t'))||p(\mathbf{A}^{thres}(t')))\]
Similarly, the anomaly coefficient of nodes can be calculated by \[\eta_{n}(t') = KL(p(\mathbf{X}(t')|\mathbf{A}(\textless t'),\mathbf{X}(\textless t'),\mathbf{Z}({\leq} t'))||p(\mathbf{X}^{thres}(t')))\]
Then the anomalous levels of different dynamic networks can be determined by $\eta_{e}(t')$ and $\eta_{n}(t')$.

\begin{table*}[t]
\centering
  \caption{Dataset statistics}
  \label{datasets}
  \begin{tabular}{c|cccl}
    \hline
    Metrics & UCI message & HEP-TH &Social evolution & Github\\
    \hline
    \hline
    Snapshots & 56 & 40 & 54 & 24\\
    Nodes & 1,899 & 7623 & 84 & 1049\\
    Edges & 688-13838 &769-34941 & 303-1172 & 4396-14238\\
    Average Density & 0.003837 & 0.00117 & 0.21740 & 0.00107\\
    Dimension of Node Attributes &- & - & 168 & 2098\\
  \hline
  \end{tabular}
\end{table*}

\section{Experiments}
In this section, we evaluate the performance of the proposed H-VGRAE and compare it to state-of-the-art anomaly detection methods on dynamic networks.
\subsection{Experimental Setup}
\subsubsection{Datasets}
\

We choose four real-world datasets of dynamic networks, UCI message \cite{opsahl2009clustering}, HEP-TH \cite{leskovec2007graph}, Social evolution \cite{trivedi2019dyrep}, and Github \cite{trivedi2019dyrep}.

\textbf{UCI message} is a directed network based on an online community of students at the University of California, Irvine.
Each node represents a user and each directed edge represents a message interaction between two users.
We aggregate all the edges in a day to form a snapshot and then discard the snapshots with few edges.
This network does not have any node attribute.

\textbf{HEP-TH} is a dataset created on the all the citations of the papers in High Energy Physics Theory conference from January 1993 to April 2003 (124 months).
If a paper $i$ cites paper $j$, the graph contains a directed edge from $i$ to $j$.
For the first 40 month, we create a citation graph using all the papers published up to that month to be a snapshot.
No node attribute is provide for this network.

\textbf{Social evolution} is collected from Jan 2008 to June 30, 2009 and released by MIT Human Dynamics Lab \cite{DeRep}.
Each node represents a user and each edge is created by Close Friendship records.
We consider the collected information from Jan 2008 until Sep 10, 2008 (i.e. survey date) to form the initial network and use cumulative data for 5 days periods of to form a snapshot of dynamic network (54 snapshots) and the remove nodes with few edges.
For this dataset, we consider Calls and SMS records between users as node attributes.

\textbf{Github} is a collected from Jan 2013 to Dec 2013.
Each node represents a Github user and each edge is set based on the Follow records.
We aggregate 15 days of records to be a snapshot and form 24 snapshots.
For this dataset, the Star and Watch records are treated as node attributes.

For UCI message and HEP-TH, which have no node attributes, we consider the $N$-dimensional identity matrix as node attributes at time $t$.

\subsubsection{Baselines}
\

%

The competing methods compared with H-VGRAE in this paper are summarized as follows.
\begin{itemize}
  \item GOutlier \cite{aggarwal2011outlier}: GOutlier detects anomalies based on a structural connectivity model. It builds a generative
model for edges in a node cluster, and the model can also be used to produce anomalous score for a given edge.
GOutlier cannot use node attributes.
  \item CM-Sketch \cite{ranshous2016scalable}: CM-Sketch defines the local and global structural feature and historical behavior near an edge to measure whether the edge is anomalous or not. It utilizes Count-Min sketch for approximating these properties.
  CM-Sketch cannot utilize node attributes.
  \item NetWalk \cite{yu2018netwalk}: NetWalk first uses a standard autoencoder to generate node embeddings from random walks, and then detects anomalies by clustering the node/edge embeddings. NetWalk cannot handle dynamic networks with node attributes.
  \item AddGraph \cite{zheng2019addgraph}: AddGraph is also a network embedding based method that designs a graph-RNN to capture ST patterns, and employ a selective negative sampling mechanism and margin loss in training of AddGraph in a semi-supervised fashion.
  AddGraph can handle dynamic networks with node attributes.
  \item VGRNN \cite{hajiramezanali2019variational}: VGRNN is a stochastic network embedding model with single scale ST features and share weights between the encoder and the decoder.
  VGRNN can utilize node attributes but only maximize the likelihood of edges.
\end{itemize}

For the models that cannot handle node attributes, only adjacency matrices are used.
Moreover, we design ablation experiments to show the effectiveness of the components of H-VGRAE.
\begin{itemize}
  \item S-VGRAE-I: S-VGRAE-I is a simple version of H-VGRAE with single ST scales ($M=1$). Due to the single layer of latent variables, the information sharing mechanism cannot work.
  \item H-VGRAE-G: H-VGRAE-G is a Gaussian version of H-VGRAE without normalizing flows, which limits the latent variable $\mathbf{Z}(t)$ in both posterior and prior to be Gaussian.
  \item H-VGRAE-I: H-VGRAE-I is a version of H-VGRAE without information sharing mechanism between the encoder and the decoder.
\end{itemize}

\subsubsection{Model Configurations}
\

%

For all datasets, we set the scale number $M=2,3$ in our standard H-VGRAE, which is denoted by H-VGRAE $(M)$.
The GNN is set to be GCN \cite{GCN}.
The output dimensions of FC, GNN, $\text{DRNN}^{(i)}$, GNN, GNN Linear and GNN Softplus are 64, 64, 64, 32, 16, and 16.
All the models are trained by the Adam optimizer with a learning rate of 0.01 for 100 epochs.
The dropout rate is 0.2 and the weight decay parameter for regularization is 1e-5.

\begin{table*}[t]

  \centering
  \caption{Anomaly detection performance comparison}
  \label{results}
  \begin{tabular}{c|ccc|ccc|ccc|ccl}
    \hline
    \multirow{2}*{Methods} &
    \multicolumn{3}{c|}{UCI message} & \multicolumn{3}{c|}{HEP-TH} & \multicolumn{3}{c|}{Social evolution} & \multicolumn{3}{c}{Github}\\
    \cline{2-13}
    & 1\% & 5\% & 10\% & 1\% & 5\% & 10\% & 1\% & 5\% & 10\% & 1\% & 5\% & 10\% \\
    \hline
GOutlier            & 0.7181 & 0.7053 & 0.6707 & 0.6964 & 0.6813 & 0.6322 & 0.6925 & 0.6844 & 0.6751 & 0.6182 & 0.6015 & 0.5842 \\
CM-Sketch           & 0.7270 & 0.7086 & 0.6861 & 0.7030 & 0.6709 & 0.6386 & 0.7086 & 0.6955 & 0.6773 & 0.6247 & 0.6066 & 0.5871 \\
NetWalk             & 0.7758 & 0.7647 & 0.7226 & 0.7489 & 0.7293 & 0.6939 & 0.7484 & 0.7369 & 0.7240 & 0.6567 & 0.6398 & 0.6243 \\
AddGraph            & 0.8083 & 0.8090 & 0.7688 & 0.7714 & 0.7573 & 0.7259 & 0.7735 & 0.7621 & 0.7536 & 0.7257 & 0.7103 & 0.6994 \\
VGRNN               & 0.8151 & 0.8149 & 0.7813 & 0.7825 & 0.7662 & 0.7318 & 0.7951 & 0.7873 & 0.7769 & 0.7557 & 0.7388 & 0.7325 \\
S-VGRAE-I           & 0.8159 & 0.8162 & 0.7827 & 0.7831 & 0.7679 & 0.7320 & 0.7966 & 0.7885 & 0.7787 & 0.7571 & 0.7413 & 0.7342 \\
H-VGRAE-G ($M=2$)   & 0.8195 & 0.8201 & 0.7875 & 0.7901 & 0.7709 & 0.7370 & 0.8097 & 0.8014 & 0.7906 & 0.7692 & 0.7509 & 0.7470 \\
H-VGRAE-I ($M=2$)   & 0.8272 & 0.8243 & 0.7894 & 0.7943 & 0.7755 & 0.7403 & 0.8118 & 0.8050 & 0.7941 & 0.7725 & 0.7536 & 0.7511 \\
H-VGRAE ($M=2$)     & 0.8350 & 0.8317 & 0.8007 & \bf{0.8057} & 0.7901 & 0.7529 & \bf{0.8173} & 0.8099 & 0.8002 & 0.7806 & 0.7624 & 0.7579 \\
H-VGRAE ($M=3$)     & \bf{0.8366} & \bf{0.8325} & \bf{0.8021} & 0.8055 & \bf{0.7926} & \bf{0.7541} & 0.8170 & \bf{0.8124} & \bf{0.8017} & \bf{0.7820} & \bf{0.7637} & \bf{0.7598} \\
  \hline
  \end{tabular}
\end{table*}

For performance comparison, we evaluate different methods based on their ability to correctly detect true and false anomalous edges.
To evaluate the robustness of the anomaly detectors, the metric is set to be area under the ROC curve (AUC) score.
In all of our experiments, we test the models on the last 10 snapshots of dynamic networks while learning the parameters of the models based on the rest of the snapshots.
In the training stage, we use 50\% edges of the training snapshots to train the models.
Due to the difficulties to obtain ground truth anomalies in the test phase, we inject 1\%, 5\% and 10\% of anomalous edges following the method in \cite{yu2018netwalk}.
All the experiments are conducted on the deep learning platform, PyTorch \cite{adam2017automatic}, on a Linux server.

\subsection{Results and Discussions}

Table \ref{results} summarizes the results for anomalous link detection on different datasets.
It is evident that the network embedding based methods (NetWalk, AddGraph, VGRNN, and H-VGRAE) outperform traditional models (GOutlier and CM-Sketch) on all the four datasets.
This illustrates the effectiveness of complex pattern captured by network embedding, compared with the artificial features in GOutlier and CM-Sketch.
In all the datasets, our H-VGRAE consistently outperforms other competitors by a large margin.

\begin{figure*}[t]
\centerline{\includegraphics[width=0.95\textwidth]{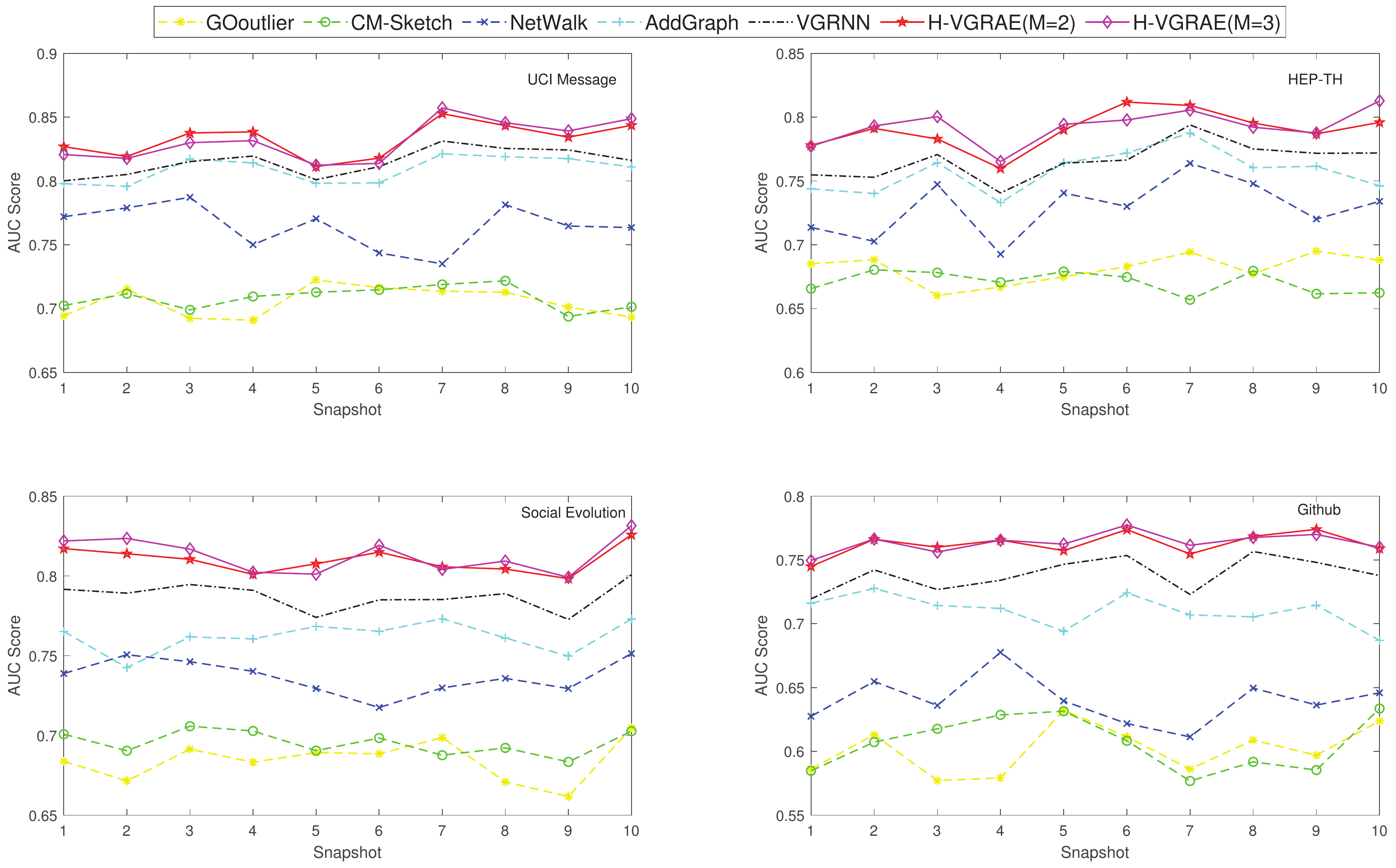}}
\caption{The AUC scores of the models in the 10 snapshots on all datasets with 5\% anomalies.}
\label{AUC}
\end{figure*}

The effectiveness of multi-scale ST features and stochasticity modeling is also shown in Table \ref{results}.
The ST features are crucial for anomaly detection in dynamic networks, which can be shown by comparing the method without extraction of ST features (NetWalk) and the methods with ability of ST modeling (AddGraph, VGRNN and our H-VGRAE).
We can see that the ST feature modeling methods have higher AUC scores.
Moreover, the robustness achieved from stochastic latent representation is also validated by comparing the deterministic methods (NetWalk and AddGraph) and the stochastic methods (VGRNN and our H-VGRAE).
This validates that random variables can carry more information than deterministic hidden states specially for capture complex temporal patterns in dynamic networks.
From the perspective of ensemble learning, the multiple sampls of latent variables essentially achieves the ensemble learning in a single model.

To show the effectiveness of multi-scale ST feature extraction, we can compare models in a two steps.
Firstly, the single scale version S-VGRAE-I outperforms VGRNN for only near 0.001 in average.
The main difference between S-VGRAE-I and VGRNN is that VGRNN shares weights between decoder and the encoder while S-VGRAE-I does not.
With more parameters in S-VGRAE-I, the outperformance is not evident, which means that it is not the number of parameters that limits the performance of VGRNN.
Secondly, the multi-scale version H-VGRAE-I ($M=2$) outperforms the single-scale version S-VGRAE-I consistently for more than 0.07, and H-VGRAE-I ($M=3$) outperforms H-VGRAE-I ($M=2$) on most scenarios.
At the two exception point, the AUC score of H-VGRAE-I ($M=3$) is lower than that of H-VGRAE-I ($M=2$) for only $2e^{-4}$ and $3e^{-4}$, which is negligible.
Therefore, we argue that enlarging the number of parameters should be combined with some expert knowledge, which is the multi-scale ST information in H-VGRAE.
The same principle is also used in CNNs by sharing weights in the convolution kernels to focus local reception field like human vision.

%
%

\begin{figure}
\centering
\subfigure[UCI Message.]{
\begin{minipage}[t]{4cm}
\centering
\includegraphics[width=4cm]{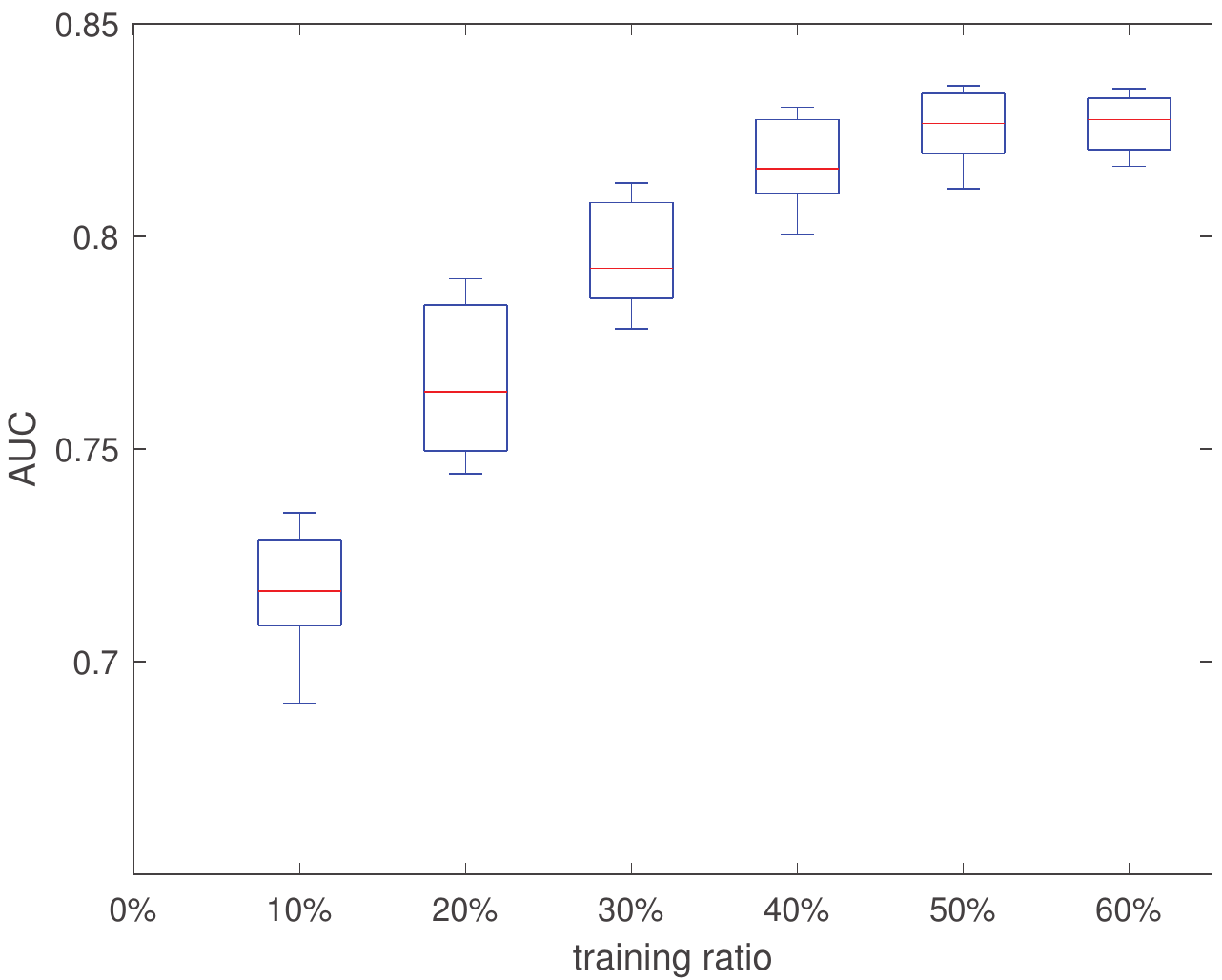}
\end{minipage}
}
\subfigure[Github.]{
\begin{minipage}[t]{4cm}
\centering
\includegraphics[width=4cm]{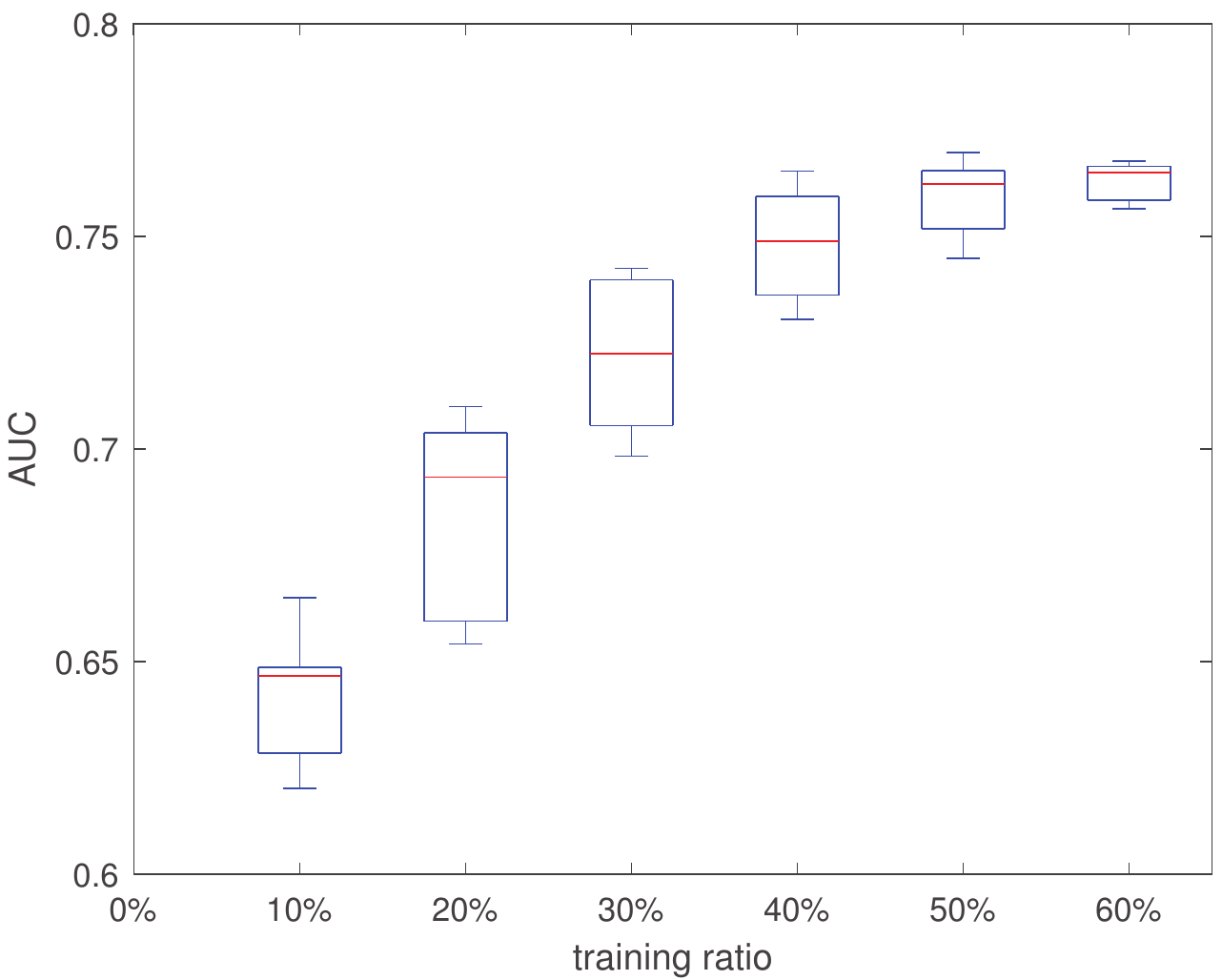}
\end{minipage}
}
\caption{The AUC score of H-VGRAE($M=2$) with different training ratios.}
\label{training_ratio}
\end{figure}

The ablation study of H-VGRAE can clearly show the effectiveness of the components in standard H-VGRAE.
Comparing H-VGRAE-G ($M=2$) with H-VGRAE ($M=2$), we can see that H-VGRAE ($M=2$) outperform H-VGRAE-G ($M=2$) for more than 0.01, which demonstrates that Gaussian latent distribution may not be the best choice for node representations.
The usefulness of the information mechanism can be shown by the outperformance of H-VGRAE ($M=2$) when comparing to H-VGRAE-I ($M=2$).

\subsection{Latent Representation Visualization}
To explain how H-VGRAE works for anomaly detection, we visualize the $\mathbf{Z}$-space representations.
Because H-VGRAE is a reconstruction based model, the latent $\mathbf{Z}$-space representations directly influence the reconstruction probabilities.
During offline model training, H-VGRAE learns the latent representations of normal behaviors of training data.
In the online detection stage, H-VGRAE also encodes the input observations according to the normal pattern.
If an input snapshot during online detection contains anomalous edges, its $\mathbf{Z}$-space representation should still concerntrate on normal parts, so the reconstruction probabilities of these anomalous edges would be low.
Due to the constraint of visualization, we set H-VGRAE($M=3$) with $d=1$ and visualize $(\mathbf{Z}^{(1)}(t),\mathbf{Z}^{(2)}(t),\mathbf{Z}^{(3)}(t))$ at each node on UCI Message dataset.

\begin{figure}[t]
\centerline{\includegraphics[width=0.48\textwidth]{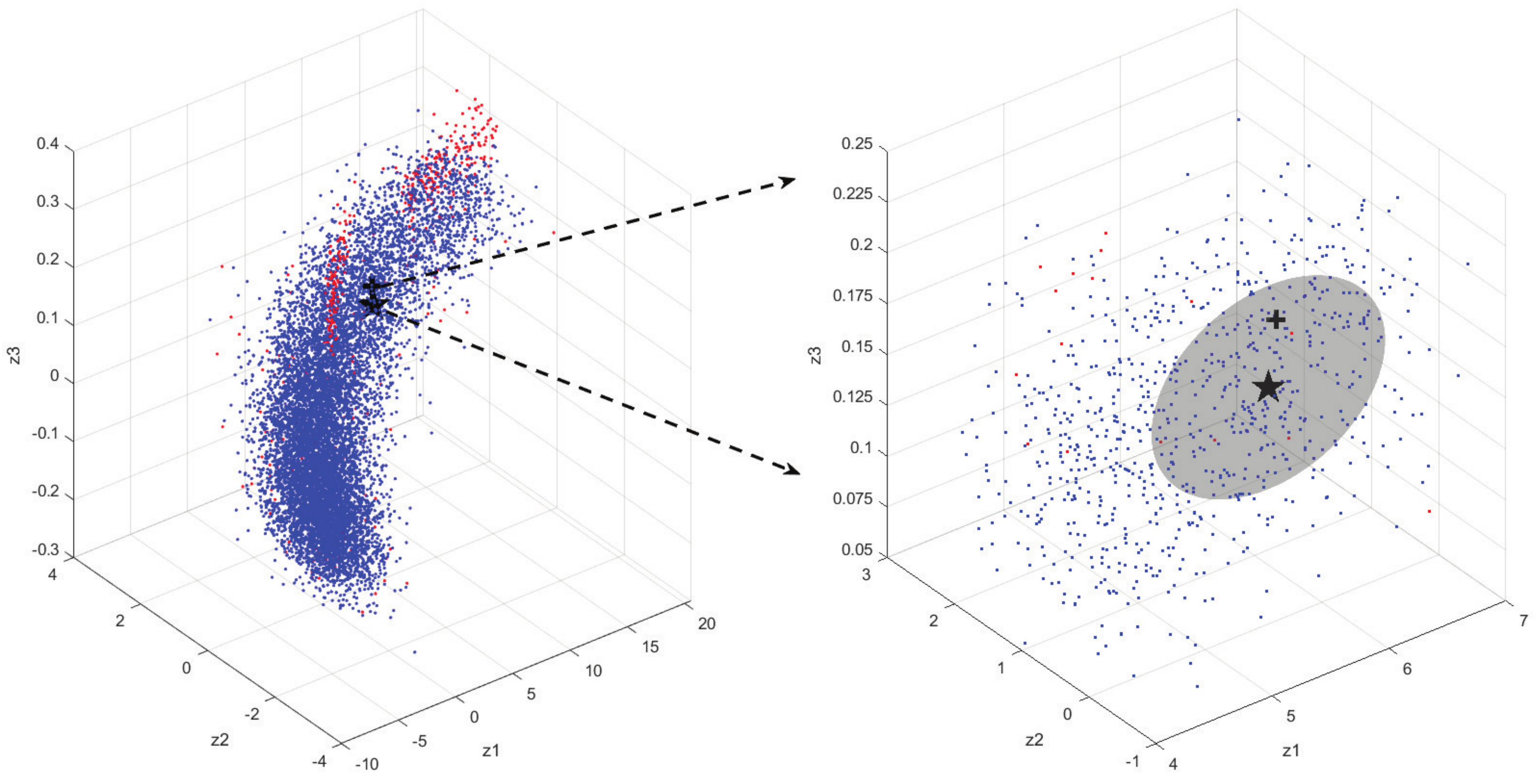}}
\caption{Visualization of the latent representation in UCI message.}
\label{zvis}
\end{figure}

Figure \ref{zvis} shows the 3-D $\mathbf{Z}$-space variables learned from the UCI message dataset by H-VGRAE($M=2$).
$(\mathbf{Z}^{(1)}(t),\mathbf{Z}^{(2)}(t),\mathbf{Z}^{(3)}(t))$ are set to their corresponding mean value.
We find that the $\mathbf{Z}$-space representation of anomalous nodes highly overlap those normal nodes, indicating that their $\mathbf{Z}$-space representations are quite similar.
This phenomenon demonstrates that H-VGRAE learns the normal behaviors well and tries to encodes the anomalous inputs in a normal way.
In such a way, the reconstruction is also based on the normal behaviors.
Then, the anomalous edges are the inputs deviated from the normal behaviors, which means these edges should not exist according to the normal pattern and the reconstruction probabilities of the anomalous edges are low.

\subsection{Parameter Sensitivity}
We examine the influence of hyper-parameters on H-VGRAE, including the number of scales $M$, the size of dimensions of latent variables $d$ and the training ratio of edges.

\begin{figure}
\centering
\subfigure[UCI Message.]{
\begin{minipage}[t]{4cm}
\centering
\includegraphics[width=4cm]{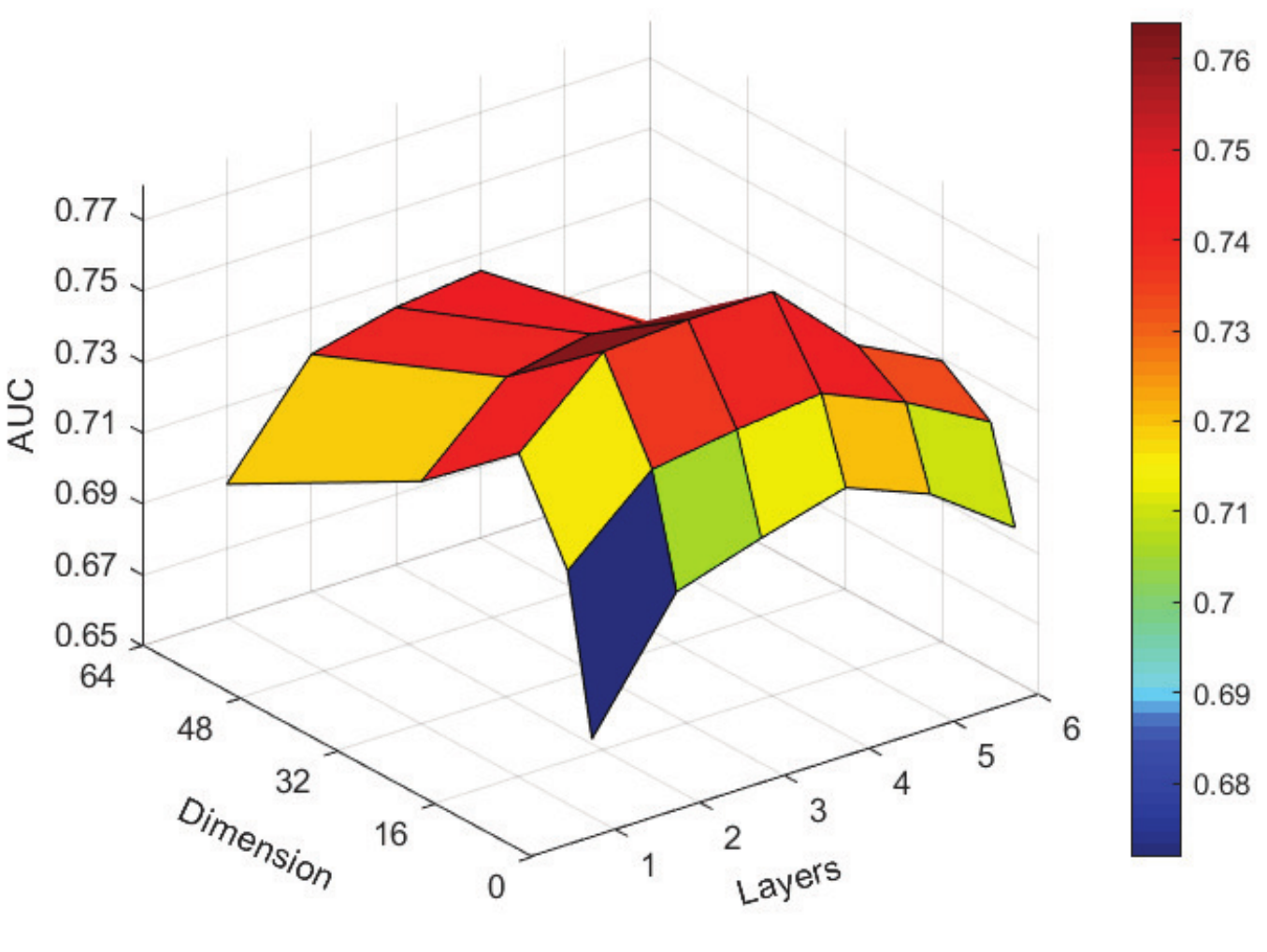}
\end{minipage}
}
\subfigure[Github.]{
\begin{minipage}[t]{4cm}
\centering
\includegraphics[width=4cm]{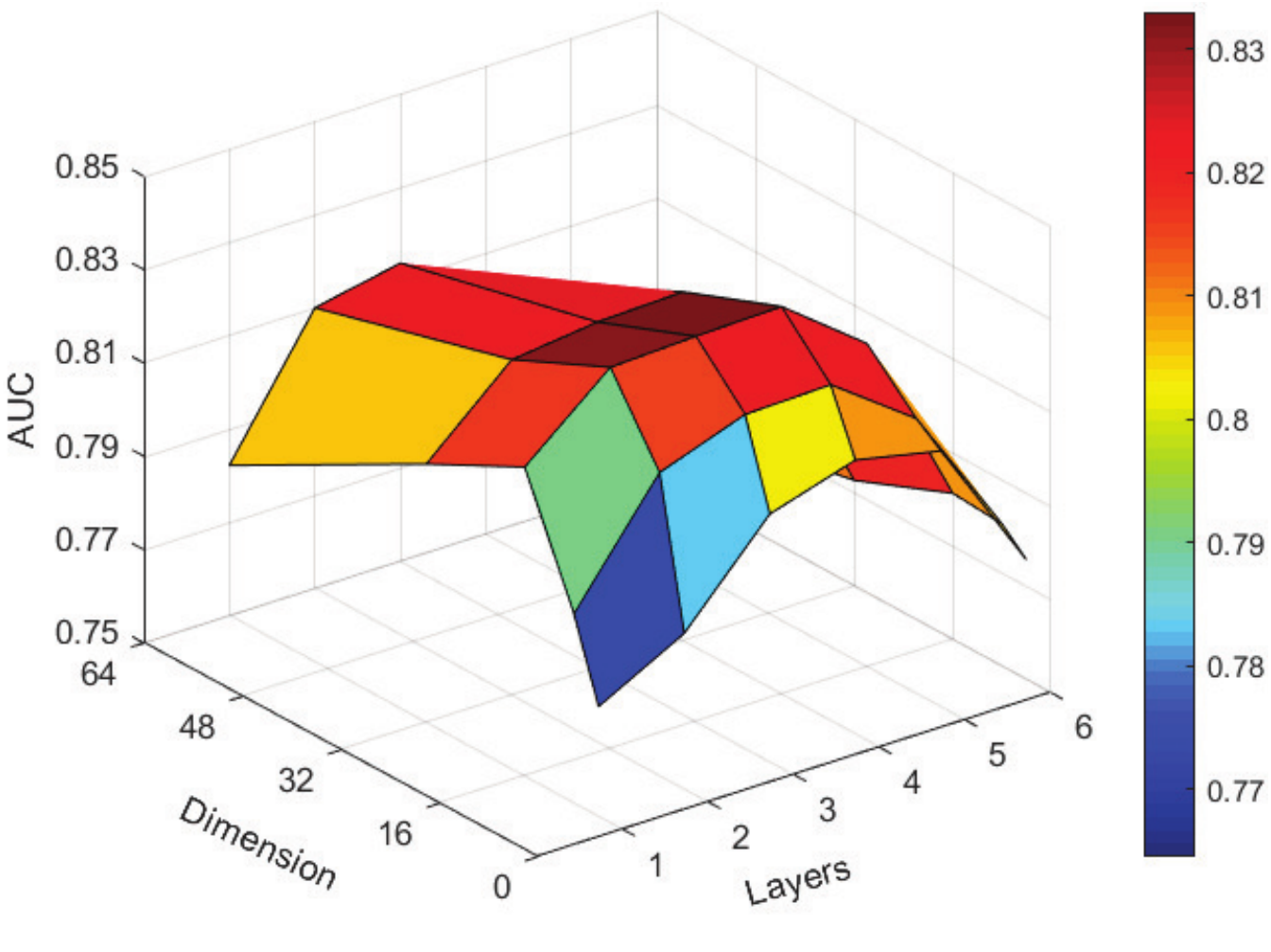}
\end{minipage}
}
\caption{The sensitivity of the scale and the latent dimension.}
\label{SEN}
\end{figure}

To evaluate the influence of $M$ and $d$, we set the range of $M$ to $\{1,2,3,4,5,6\}$ and the range of $d$ to $\{4,8,16,32,64\}$.
Other parameters are set to default.
The performance of H-VGRAE is examined on the anomaly detection tasks on the datasets of UCI Message and Github with 5\% anomalies.
The results are summarized in Figure \ref{SEN}.
The AUC score of H-VGRAE increases significantly from $M=1$ to $M=2$.
Further improvement is not evident from $M=2$ to $M=4$, and the performance degrades after $M=5$.
Due to the exponentially increase of dilation factor, $M=5$ means the skip connection would be $16$ timestamps, which is larger than the test snapshots $10$.
Thus the upper-scale $\text{DRNN}(\cdot)$ in $M=5,6$ would degrade to FC layers and loss the ability of modeling temporal dependency.
Moreover, with the increase of the number GNN layers, the node representations would be smooth and ends to the same if stacking too many GNN layers \cite{dehmamy2019understanding}.
So the upper scales would be noise to interrupt the inference.

With the increase of $d$, the AUC score of H-VGRAE increase gradually from $d=4$ to $d=16$, and then drops after $d=32$.
The reason is that, when $d$ is small, useful information may be missing; while when $d$ is too high, there can be some noise captured by the latent representation.
The dimension of the previous layer is $32$.
When $d\textless 32$, the overall structure of H-VGRAE would be of the “sandwich” structure to help the model sufficiently apply bottleneck strategy \cite{Sze2017Efficient}.
The structure can perform the mechanism similar with dimension reduction to preserve critical information.

Then, we evaluete the influence of the training ratio of the edges.
The range of the training ratio is set to $\{10\%,20\%,30\%,40\%,50\%,60\%\}$ and other parameters are set to optimum.
We use the UCI message dataset and Github dataset with 5\% anomalies in test data, and record the AUC scores of the 10 snapshots in the test stage.
As shown in Figure \ref{training_ratio}, with the increase of the training ratio from 10\% to 40\%, the AUC scores increase sharply, and then the performance stays relatively stable.

\section{Conclusion}
We have presented a robust semi-supervised model, H-VGRAE, to detect anomalies in dynamic networks by learning hierarchical stochastic network embeddings from multi-scale spatial-temporal features.
The H-VGRAE can handle scenarios with time-varying edges and node attributes.
To break the limits of Gaussian assumption and avoid the KL vanishing problem, the normalizing flow technique and the predictive prior mechanism is employed.
The difficulty of optimizing deep hierarchies of stochastic conditional layers is alleviated by the designed information sharing mechanism between the conditional random variables in the encoder and the decoder.
Based on the reconstruction probabilities, the anomalous edges and nodes can be detected with interpretable latent representations.
Through extensive experiments, the proposed H-VGRAE outperforms state-of-the-art approaches on four real-world datasets of dynamic networks.

In the future, we will improve H-VGRAE by exploring feature extractors that can tackle the problem of feature degradation with the increase of scales.
For the dynamic networks with time-varying number of nodes, new mechanisms of extracting temporal features should be designed to handle variable dimensions of hidden states.

\bibliographystyle{IEEEtran}
\bibliography{ref.bib}

\end{document}